\documentclass[11pt]{article}
\usepackage{jheppub}

\input{epsf}
\usepackage{epsfig}
\usepackage{amssymb}
\usepackage{amsfonts}
\usepackage{amsbsy}
\usepackage[all]{xy}
\usepackage{amsmath}
\usepackage{dsfont}
\usepackage{bbm}
\usepackage{fixmath}
\usepackage{upgreek}
\usepackage{amssymb,amscd}
\usepackage{mathrsfs}
\usepackage{amsmath,amsthm}
\usepackage{slashed}
\usepackage{dsfont}
\usepackage{tikz}
\usepackage[utf8]{inputenc}
\usetikzlibrary{positioning}
\usetikzlibrary{shapes.geometric, arrows}
\tikzstyle{rect} = [rectangle,rounded corners,minimum width=3cm, minimum height=1cm, text centered, draw=black]
\tikzstyle{arrow} = [thick,->,>=stealth]
\usepackage{stackrel}

\tikzset{
  font={\fontsize{8pt}{12}\selectfont}}

\makeatletter
\DeclareFontFamily{OMX}{MnSymbolE}{}
\DeclareSymbolFont{MnLargeSymbols}{OMX}{MnSymbolE}{m}{n}
\SetSymbolFont{MnLargeSymbols}{bold}{OMX}{MnSymbolE}{b}{n}
\DeclareFontShape{OMX}{MnSymbolE}{m}{n}{
    <-6>  MnSymbolE5
   <6-7>  MnSymbolE6
   <7-8>  MnSymbolE7
   <8-9>  MnSymbolE8
   <9-10> MnSymbolE9
  <10-12> MnSymbolE10
  <12->   MnSymbolE12
}{}
\DeclareFontShape{OMX}{MnSymbolE}{b}{n}{
    <-6>  MnSymbolE-Bold5
   <6-7>  MnSymbolE-Bold6
   <7-8>  MnSymbolE-Bold7
   <8-9>  MnSymbolE-Bold8
   <9-10> MnSymbolE-Bold9
  <10-12> MnSymbolE-Bold10
  <12->   MnSymbolE-Bold12
}{}

\let\llangle\@undefined
\let\rrangle\@undefined
\DeclareMathDelimiter{\llangle}{\mathopen}%
                     {MnLargeSymbols}{'164}{MnLargeSymbols}{'164}
\DeclareMathDelimiter{\rrangle}{\mathclose}%
                     {MnLargeSymbols}{'171}{MnLargeSymbols}{'171}
\makeatother

\def\be{ \begin{equation} }
\def\ee{ \end{equation}}

\def\log{{\rm log}}

\def\half{\frac{1}{2}}

\def\one{{\hbox{ 1\kern-.8mm l}}}

\def\CH {{\cal H}}

\def\CM {{\cal M}}
\def\CN {{\cal N}}
\def\CO {{\cal O}}

\def\CW {{\cal W}}

\def\CO {{\cal O}}

\def\CH {{\cal H}}

\def\CQ {{\cal Q}}
\def\CS {{\cal S}}
\def\CT {{\cal T}}
\def\CU {{\cal U}}

\def\IC{\mathbb{C}}

\def\IP{\mathbb{P}}

\def\IR{{\mathbb{R}}}

\def\IZ{{\mathbb{Z}}}

\def\fg{\mathfrak{g}}

\def\rmk#1{\bigskip\noindent{\bf Remarks} }
\def\cnj#1{\bigskip\noindent{\bf Conjecture:} }

\DeclareMathAlphabet{\mathpzc}{OT1}{pzc}{m}{it}

\def\half{\frac{1}{2}}

\title{The String Landscape, the Swampland, and the Missing Corner}

\author[a]{T. Daniel Brennan}
\author[b]{Federico Carta\footnote{La Caixa-Severo Ochoa Scholar}} 
\author[c]{Cumrun Vafa}
\affiliation[a]{NHETC and
Department of Physics and Astronomy, Rutgers University \\
126 Frelinghuysen Rd., Piscataway NJ 08855, USA}
\affiliation[b]{Instituto de F\'isica Te\'orica UAM-CSIC,\\
Universidad Aut\'onoma de Madrid, Cantoblanco, 28049 Madrid, Spain}
\affiliation[c]{Jefferson Physical Laboratory, Harvard University\\
Cambridge, MA 02138, USA}

\abstract{  We give a brief overview of the string landscape  and techniques used to construct string compactifications.  We then explain how this motivates the notion of the swampland and review a number of conjectures that attempt to characterize theories in the swampland.  
We also compare holography in the context of superstrings with the similar, but much simpler case of topological string theory.  For topological strings, there is a direct definition of topological gravity based on a sum over a ``quantum gravitational foam.''  In this context, holography is the statement of an identification between a gravity and gauge theory, both of which are defined independently of one another.  This points to a missing corner in string dualities which suggests the search for a direct definition of quantum theory of gravity rather than relying on its strongly coupled holographic dual as an adequate substitute (Based on TASI 2017 lectures given by C. Vafa). }

\preprint{IFT-UAM/CSIC-17-105}

\begin{document}

\maketitle

These lecture notes from TASI 2017 give a brief overview of some of the open problems in string theory. We will be generally motivated by the philosophy that string theory is ultimately supposed to describe the fundamental laws of our universe. String theory is so versatile that it can be used to study a wide array of physical problems such as various topics in condensed matter and quark-gluon plasma or aspects of quantum fields theories in diverse dimensions.   Much of the recent work using string theory has been focused on using its properties to solve specific problems rather than developing our understanding of string theory as a fundamental description of our universe.   Here we aim to discuss topics which we hope will be useful in bringing string theory closer to observable aspects of fundamental physics.

With this philosophy in mind, we will begin these lectures by reviewing some of what we know about string theory and its possible application to the universe by describing some generalities about the space of low energy theories theories coming from string theory compactifications: this is called the ``string landscape.''   Supersymmetry plays a key organizing principle in this context.
This will naturally lead us to investigate the question of how we know a priori if a low energy theory is in the landscape or it is not.   The set of low energy physics models which look consistent but ultimately are not when coupled to gravity, is called the ``swampland.''  Finding simple criteria to distinguish the swampland from the landscape is of great importance.  In particular such criteria can lead to concrete predictions for our universe as we will discuss later.
We review a number of conjectures which are aimed at distinguishing the swampland from the landscape.

The string landscape and the swampland will be the topic of the first two lectures. In the third lecture, which is on a somewhat disjoint topic, we review critically where we are in our current understanding of quantum gravity.  In this lecture we use the toy example of topological string theory, for which much more is known,
 to shed a new light on what shortcomings we have in our current formulation of quantum gravity from string theory.  In topological string theory the gravitational theory can be formulated in terms of a non-commutative $U(1)$ gauge theory whose configurations can be interpreted as defining a quantum gravitational foam.  This is holographically dual to a Chern-Simons theory.  The equivalence of these two theories is the content of holography for topological strings. 
  In the case of full string theory, a direct definition of quantum gravity is missing and holography is viewed as a substitute definition.  The analogy with topological strings suggests that there is a missing corner in string dualities and that we should try to find a direct definition of quantum gravity.

\section{Lecture 1: The String Landscape}
\label{sec:Lec1}

\subsection{Review of String Theories}

In the late 1980's it was realized that there were five perturbatively consistent, distinct ten-dimensional string theories.  Early efforts to find the four-dimensional standard model with gravity as a low energy limit of string theory focused on compactifying those theories on various manifolds. Initially, the Heterotic string theories were considered the most promising theories to produce the standard model because of their inherent non-abelian gauge symmetries
\footnote{Recall that D-branes were discovered only in the late 80's - early '90s, and before this moment it was largely unknown how to realize gauge groups in Type II theories \cite{Dai:1989ua}.}. In  mid 1990's, it was realized that these five theories were all related by different dualities and non-perturbative completions, thus launching the ``duality era'' \cite{Witten:1995ex}. 

The five original types of superstring theories are Type I, Type IIA, Type IIB, Heterotic $E8\times E8$ (HE), and Heterotic $SO(32)$ (HO)\footnote{The gauge group is actually $Spin(32)/\IZ_2$, but it is convention to say $SO(32)$ since they have isomorphic Lie algebras \cite{Gross:1984Dd,Gross:1985fr,Gross:1985rr}.}. There is also a conjectured 11D theory called M-theory which is not a string theory and has a low energy effective action given by 11D $\CN=1$ supergravity. This theory is highest dimensional supergravity theory.  The Type IIA has a $\CN=(1,1)$ supersymmetry  and thus is non-chiral.  Type IIB has $\CN=(2,0)$ and is thus chiral.  Heterotic and Type I strings are both chiral carrying $\CN=(1,0)$ supersymmetry\footnote{$\CN=(p,q)$ indicates that there are $p-$SUSY operators in the left-handed spin representation of the d-dimensional Lorentz group and $q-$SUSY operators in the right-handed spin representation.  More precisely, in even dimensions, the spin representations of the Lorentz group are distinguised by chirality which gives the left- and right-handed spin representations. However, in 4D and 8D, these representations are related by complex conjugation and hence are identical so that we only have this chiral notation for 2D, 6D, and 10D. }.

 The duality web suggests that all these theories are related by dualities.  Let us denote  $k$-dimensional Minkowski space by $M_k$.  The dualities are generated by the following equivalences (also see Figure \ref{fig:dualities}):
\begin{itemize}
\item Type I is related to HO by S-duality (inversion of coupling constant),
\item HO is related to HE on $M_9\times S^1$ with Wilson lines turned on $S^1$ \cite{Narain:1986am} ,
\item Type IIA on $M_{10}$ is equivalent to M-theory on $M_{10}\times S^1$ where the radius $R$ of $S^1$ is related to Type IIA coupling constant by $R^3=\lambda^2$,
\item Type IIA on $M_9\times S^1_R$ is related to Type IIB on $M_9\times S^1_{R'}$ by T-duality where $R'=\ell_s^2/R$ where $\ell_s$ is the string length, 
\item M-theory on $M_{10}\times S^1/\IZ_2$ is related to HE on $M_{10}$.  Physically, M-theory on this interval (whose length is related to Heterotic string coupling constant) has M9-branes on each end of the interval each of which carry an $E_8$ gauge group which together give rise to the 10D HE theory \cite{Horava:1996ma,Moore:2000fs}. 
\end{itemize}

\tikzstyle{block} = [rectangle, draw, fill=blue!20, 
    text centered, rounded corners, minimum height=1.5em]
\tikzstyle{line} = [draw, -latex']
\tikzstyle{cloud} = [draw, ellipse,fill=red!20, node distance=6cm,
    minimum height=2em]
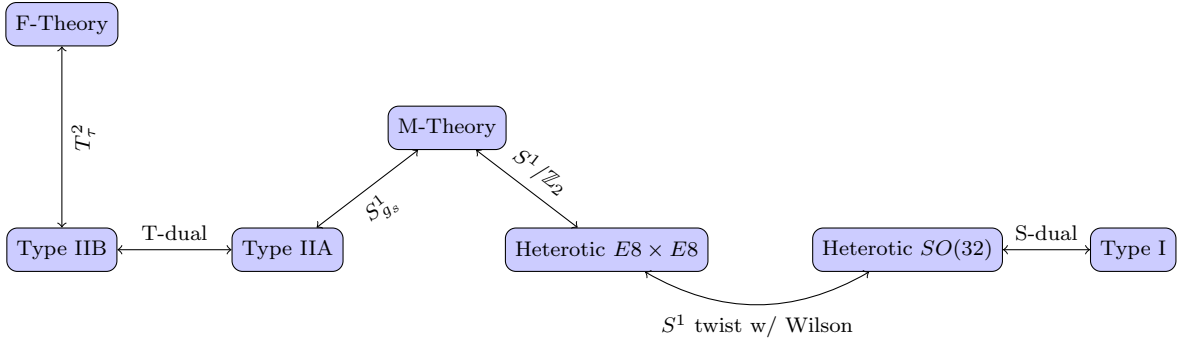
\begin{figure}
\begin{tikzpicture}[node distance = 3cm, auto,<->]
\node [block] (mtheory) {M-Theory};
\node (IIA) [block,below left of = mtheory,yshift=+0.5cm] {Type IIA};
\node (IIB) [block,left of =IIA] {Type IIB};
\node (F) [block,above of= IIB] {F-Theory};
\node (HE) [block,below right of=mtheory,yshift=+0.5cm] {Heterotic $E8\times E8$};
\node (HO) [block,right of=HE,xshift=+1cm] {Heterotic $SO(32)$};
\node (I) [block, right of=HO] {Type I};
\draw[<->] (mtheory) --  (IIA) node[midway,sloped,below]{
$S^1_{g_s}$};
\draw[<->] (IIA) -- (IIB) node[midway,sloped,above]{T-dual};
\draw[<->]   (IIB) -- (F) node[midway,sloped,below]{$T^2_\tau$};
\draw[<->] (mtheory) --  (HE) node[midway,sloped,above]{$S^1/\IZ_2$};
  \path  (HE) edge[bend right] node [below] {$S^1$ twist w/ Wilson} (HO);
\draw[<->]  (HO) --  (I) node[midway,sloped,above]{S-dual};

\end{tikzpicture}
\caption{This graph shows the duality web relating all of the different descriptions of string theory. Here the vertical placement corresponds to dimension. }
\label{fig:dualities}
\end{figure}

While M-theory appears to be the most central in web of dualities between all of the different theories, in some sense, it doesn't see Type IIB string theory, even though they are clearly dual. Specifically if we compactify M-theory on $T^2$ and shrink it to zero size, we would expect to get a 9D effective theory. However, string duality tells us that we actually get Type IIB which has 10D Lorentz symmetry. This indicates that M-theory is perhaps not a very good way to describe Type IIB. 

Let us now consider Type IIB string theory. It has a complex coupling constant, $\tau$ which is $SL(2,\IZ)$ equivariant since Type IIB is self-dual under S-duality. Because of this property, we can think of $\tau$ as the complex structure of a compactification torus of some 12D theory. This 12D theory is what is referred to as F-theory \cite{Vafa:1996xn}.  
  See \cite{Denef:2008wq,Heckman:2010bq} for a review of some aspects of F-theory.

\subsection{String Compactification}

We can now try to produce lower dimensional theories and in particular a four dimensional theory that describes our universe by compactifying one of these theories to lower dimensions. By this we mean that we will take a spacetime of the form $M_d\times K$ where $K$ is some compact manifold with appropriate dimension with some characteristic size $R_K$. Then we will take the limit of $\ell_s<<R_K$ (or $l_{Planck}<<R_K$ for M-theory) and consider the low energy dynamics with energy $E<<1/R_K$, leading to an effective theory on $M_d$. Note that we do not take the limit of $R_K\to 0$ as this will lead to the introduction of light matter coming from the winding states which typically leads to a higher dimensional dual description.

Since there are so many (possibly infinitely many) manifolds we can compactify on to get lower dimensional physics, it is clear that we need some kind of organizing principle for these theories. As physicists we generally try to use symmetries to characterize different systems. As such, we will use Lorentz symmetries again for this purpose. For the time being, we will take all directions which are not part of the compact direction to be $M_d=\IR^{1,d-1}$ which has $d$-dimensional Lorentz symmetry. 

Additionally, as it is useful in classifying the string theories, it also makes sense to classify our theories by the amount of supersymmetry. For our purposes rather than using the number of supersymmetries $\CN$ which is adapted for each dimension, we will make use of the notation $N_{SUSY}$ which counts the total number of conserved supercharges, as this will not change upon flat toroidal compactifications.   As with the 10D string theories, we will also classify by the chirality of the supersymmetries. For example, Type IIA string theory has $N_{SUSY}=32$ charges and is non-chiral since it has $\CN=(1,1)$ SUSY whereas Type IIB string theory has $N_{SUSY}=32$ and is chiral since it has $\CN=(2,0)$. The relation between the two notations is that in even dimensional spacetime, a theory with $\CN=(p,q)$ has $N_{SUSY}=16(p+q)$, $N_{SUSY}=8(p+q)$, or $N_{SUSY}=(p+q)$ conserved supercharges in 10D, 6D, or 2D respectively.  Apart from the chirality information in these dimensions, $N_{SUSY}$ completely captures the structure of the super Lorentz group.

Thus far, we have been implicitly assuming that we are compactifying with compact manifolds. However, even though the name and procedure of compactification requires a compact manifold
, there has actually been a significant amount of work on ``compactifying on non-compact manifolds.'' This means that we can also study string theory on $M_d\times K$ 
 where the volume of $K$ is infinite and the theory has local normalizable modes in $M_d$ which are decoupled from gravity. 
 This is based on the way that the gravity appears in the low energy effective action for the ``compactified''  theory. 
 
 Consider the background to be $M_d\times K$ where $K$ is our compactification manifold. To leading order the gravitational part of the action on $M_d$ will be of the form
\be
S\sim Vol(K)~\int_{M_d}d^d x\sqrt{-g}~R_d~,
\ee
where $R_d$ is the $d$-dimensional Ricci scalar. Comparing to the Einstein-Hilbert action 
\be
S_{EH}=\frac{1}{16\pi G_N}\int_{\IR^d}d^dx \sqrt{-g}~R_d=M_{pl}^{d-2}\int_{\IR^d}d^dx\sqrt{-g}~R_d~,
\ee
means that $Vol(K)$ should be identified with $M_{pl}^{d-2}$. Therefore in the limit\footnote{Here we assume that we are compactifying to $d>2$.} $Vol(K)\to \infty$, $M_{pl}\to \infty$ and hence $G_N\to 0$, decoupling gravity\footnote{More generally, the procedure to decouple from gravity is more involved than just taking $Vol(K)\to \infty$. However, we will ignore this subtlety for simplicity. In general, we cannot always decouple a theory from gravity. For example when the compactified theory is not asymptotically free. It is sufficient for a theory to be asymptotically free/conformal to be able to decouple from gravity. See \cite{Seiberg:1996vs} for some additional details of decoupling gravity.}.  However depending on the structure of $K$ we may get normalizable modes of some of the fields with a finite kinetic term in $d$-dimensions.  These can in principle lead to interesting interacting quantum systems in $d$-dimensions which are not coupled to gravity.  From this, it is clear that studying non-compact ``compactifications'' is useful for studying QFTs that are not coupled to gravity. 

Non-compact backgrounds are also interesting for the purpose of studying holography. Consider string theory in $AdS$ space. This has negative curvature that requires background fluxes to be stable. In general this requires string theory to be on a space $AdS_d\times K$ where $K$ is an appropriate dimensional manifold with positive curvature to balance the total curvature in Einstein's equation.  Holography tells us that sometimes we can relate non-compact compactifications to a true compactification with $AdS$ spacetime. For example, consider Type IIB with a stack of $N$ D3-branes in $\IR^{1,9}$. Holography tells us that this is dual to $AdS_5\times S^5$ supported by $N$ units of flux. From the non-compact geometry we have ended up with a compact one! This happens by ``zooming'' in to the near horizon limit in which the transverse direction to the D3-branes looks like $\IR_+\times S^5$. In a sense the D3-brane worldvolume direction, $\IR^{1,3}$, absorbs the non-compactness of the transverse $\IR_+$ direction to become $AdS_5$, resulting in a compact geometry with flux. This is the statement of holography - that branes' interaction with gravity leaves an imprint on space which leads to an equivalence between a QFT living on the branes and its gravitational imprint where the brane has been replaced by flux 
 \cite{Maldacena:1997re}.
The distinguished role of $AdS$ space in holography begs the question if de Sitter spacetime also has similar properties. However these spaces may be impossible to obtain in string theory as we will discuss in Section \ref{sec:4D}.   For the rest of the discussion here we will focus on compactification to Minkowski background.

Thus far, the only known stable compactifications to Minkowski space require some amount of preserved supersymmetry. 
Therefore, an important part of understanding what theories are produced by compactifying string theory is understanding what conditions are sufficient and necessary to have supersymmetry for the low energy theory.  This is determined by the number of covariantly constant spinors on the compactification manifold. These are globally defined spinors which are invariant under translation along the compact manifold - they are intuitively spinors that are independent 
of their position on the internal manifold.  In many cases the statement of existence of covariantly constant spinors can be translated into a statement about the holonomy of the tangent bundle. 

This can be described as follows. Pick a tangent vector at any point on the (oriented) manifold $K$. As we parallel transport this vector along any closed path in the manifold, the vector is ``pushed around'' so that when it returns to the basepoint, the change of the vector can be described by the action of an element of $SO(n)$ for a $n$-dimensional manifold. Generically this action lifts to the spin bundle by spin representations induced from the lift of $SO(n)$ to  $Spin(n)$. However, there are special cases when this group is reduced to a subgroup\footnote{Note in order to have the reduction of the holonomy group, we must have that this holds for all choices of basepoint.} $G\subset SO(n)$. In this case, we can find an element of the spin bundle which is fixed by the lift of the action of $G$ and hence can be extended to all of $K$ by geodesic translation. 

A special class of manifold which has a guaranteed reduction of structure group are Calabi-Yau manifolds. These are K\"ahler, Ricci flat, complex manifold of real dimension $2n$ whose holonomy is given by $SU(n)\subset SO(2n)$. As we will see, these will provide an important class of internal manifold for compactification. 

Some general classes of compactification manifolds that are useful for string compactifications to 4D which preserve some amount of SUSY are:
\begin{itemize}
\item $T^n$ - the n-torus. This preserves full SUSY since the holonomy group is trivial.
\item $K3$ surface. This is the first non-trivial Calabi-Yau manifold which is a complex 2-fold (4-real dimensional). In this case the holonomy is reduced from $SO(4)\to SU(2)$. So there is a trivial direction in the spinor bundle and hence a covariantly constant spinor. This preserves $\half$-SUSY.
\item Calabi-Yau 3-manifolds ($CY^3$) are complex 3-folds with SU(3) holonomy. These preserve $\frac{1}{4}$-SUSY.
\item Calabi-Yau 4-manifolds (CY4) are complex 4-folds dimensional with SU(4) holonomy. These preserve $\frac{1}{8}$-SUSY.
\item $G2$-manifolds are 7-real dimensional manifolds with $G2\subset SO(7)$ holonomy. These preserve $\frac{1}{8}$-SUSY.
\item $Spin(7)$-manifolds are 8-real dimensional with $Spin(7)\subset Spin(8)\sim SO(8)$ holonomy. These preserve $\frac{1}{16}$-SUSY.
\end{itemize}

\subsection{Dualities of Compactified Theories}

Since we have seen that the full string theories are all interrelated by a sequence of dualities, one would expect that their compactifications are also related by dualities. As it turns out, these relations are so abundant that we can make the following observation: \\

\textbf{``Conjecture'':} Whenever the dimension, number of preserved supercharges, and chiralities of two different compactifications of string theory match, there are choices of compactification geometries such that they are dual descriptions of the same physical theory. \\

Surprisingly, we are aware of no known counter examples. In this sense, dualities in lower dimensional theories are not hard to find, but rather are hard to prevent!  One rationale for the existence of dualities is  as Sergio Cecotti puts it, ``the scarcity of rich structures''.   In particular the very existence of quantum systems of gravity is hard to arrange and if we succeed to get more than one theory with a given symmetry, there is a good chance we have landed on the same theory.

We will now briefly review some of interconnected web of string dualities with definite $N_{SUSY}$ in various dimensions.   We will not be exhaustive in the discussion below, but aim to illustrate some key examples.

\subsubsection{Adiabatic Principle}

A general principle which helps us in identifying dualities upon compactification is to use known dualities in higher dimensions to build new ones in lower dimensions.  This is based by use of a sort of ``adiabatic procedure" \cite{Vafa:1995gm} where we step-by-step reduce the dimension by acting on both sides of a given duality by a geometric procedure such as compactification.   However, in most applications the fibration data does not vary adiabatically.  Therefore, there is no rigorous reason why this should work since it is not really adiabatic in any sense of the word as things change drastically at every step. However, amazingly it does work in all known examples as long as some supersymmetry is preserved.

\subsubsection{$N_{SUSY}=32$}  
Let us start with $N_{SUSY}=32$.  M-theory in 11 dimensions, F-theory in 12 dimensions, and Type IIA and Type IIB and all their toroidal compactifications to lower dimensions have $N_{SUSY}=32$.
Moreover, except for Type IIB in 10 dimensions whose chirality is $\CN=(2,0)$ and F-theory, all these theories are non-chiral and so one would expect based on our general conjecture that they are all equivalent once compactified to the same dimension.  And further, we expect (correctly) that both the chiral and non-chiral theories to be equivalent when toroidally compactified to the same dimension. 

\subsubsection{$N_{SUSY}=16$}
Consider the case of $N_{SUSY}=16$.
This includes Heterotic and Type I theories in 10D as well as M-theory on $S^1/\IZ_2$.  We have already discussed dualities among these, so let us move on to further dualities with this much supersymmetry.
 We will start by compactifying F-theory. Recall that in order to make sense of F-theory we need a torus embedded in our compactification manifold - that is we need at least some kind of torus fibration. We will begin by compactifying F-theory on an elliptic K3-manifold. This may be written as a torus fibration over $\IP^1$
\be
\vcenter{\vbox{\xymatrix{T^2\ar[r]&K3\ar[d]\\&\IP^1
}}}\qquad K3=\{y^2=x^3+f(z)x+g(z)~|~z\in \IP^1\}~.
\ee
Note that this reduces the $N_{SUSY}=32$ of F-theory to an 8d, $N_{SUSY}=16$ theory. The Heterotic and Type I theories also have $N_{SUSY}=16$, but they are 10D theories. So if we reduce them on a $T^2$, we get an 8d theory with $N_{SUSY}=16$. As it turns out, these theories are dual to each other \cite{Vafa:1996xn}. 

One can check aspects of this duality by studying singularity structure of the K3 manifold. The key to the duality is that the singularities of K3 manifold, which are points where the $T^2$ fiber degenerates\footnote{These singularities have a classification given by Kodaira, which mirrors the more well known one for an ADE classification of $K3$ singularities. See Section \ref{sec:McKay} for  a discussion of ADE singularities.}, which are interpreted in Type IIB setup as giving rise to a system of $(p,q)$ 7-branes which can lead to non-abelian gauge symmetry as expected for Heterotic or Type I toroidal compactifications. It turns out that compactification on an elliptic K3 manifold leads to singularities associated to a Lie algebra $\fg$ such that $rnk ~\fg\leq20$ and in paritcular we cannot obtain arbitrarily large rank non-abelian gauge groups in this way. 
This is compatible with the Heterotic and Type I compactifitions on $T^2$ which leads to a rank 16 plus an additional 4 coming from winding and momentum charges on each circle but we can have gauge symmetry enhancement at certain points in the moduli space which give rise to (semi-simple) factors of non-abelian groups with up to $rnk~\fg=18$ gauge groups \cite{Aspinwall:1995zi}\footnote{As it turns out this is a special case of a more general string duality which relates Heterotic string theory on a complex $n-1$ dimensional elliptically fibered Calabi-Yau $\pi_{H}:Z\to B$ to F-theory compactified on an complex $n$-dimensional Calabi-Yau $\pi_F:X\to B$ with elliptic K3-fibers over the same base. The physical motivation for this is the same as compactification on the K3 fibers again give rise to a system of 7-branes which has non-abelian gauge symmetry matching that of the Heterotic string compactified on the $T^2$ fibers with Wilson lines. These can further be shown to have the same moduli and low energy spectrum. See \cite{Donagi:1998vw} for more details.}.



Additionally, there is an important family of dualities relating F-theory and M-theory. Recall that F-theory and M-theory have the same $N_{SUSY}=32$ and hence 
 F-theory on $T^2\times S^1$ is equivalent to M-theory on $T^2$
. If we compactify F-theory on $K^{ell}\rtimes S^1$, it will have the same dimension and $N_{SUSY}$ as M-theory on $K^{ell}$ where $K^{ell}$ is an elliptically fibered manifold. Therefore, we expect these theories to be dual. This can be derived by the adiabatic principle, where we adiabatically fiber the $T^2$ factor on both sides of the 9D duality over a base manifold to construct $K^{ell}$. 


Similarly as we discussed M-theory on $S^1$ is equivalent to Type IIA. This gives rise to the chain of dualities  \cite{Vafa:1996xn}:

\begin{itemize}
\item F-theory on $K^{ell}\times S^1$ is dual to M-theory on $K^{ell}$
\item F-theory on $K^{ell}\times S^1\times S^1$ is dual to M-theory on $K^{ell}\times S^1$ and to Type IIA on $K^{ell}$
\end{itemize}

If we apply this to the case where $K^{ell}=K3$, then we find that F-theory on $K3\times S^1$, which is conjectured to be the same as Heterotic on $T^2\times S^1=T^3$, is dual to M-theory on $K3$ \cite{Witten:1995ex}.  Moreover, going one dimension down on another $S^1$ shows that Type IIA on $K3$ is dual to is dual to Heterotic on $T^4$. Generically in the compactification of Heterotic string theory on $T^4$ includes non-trivial Wilson lines along the different $S^1$ factors in the $T^4$ which break gauge symmetry. But by turning these off, our compactified theory will exhibit non-abelian gauge symmetry. This is reflected in the Type IIA side by taking the limit where non-trivial 2-cycles in the $K3$ manifold become coincident and develop orbifold singularities as above. We will comment more on this in the next section.

\subsubsection{$N_{SUSY}=8$}
Now consider the case of $N_{SUSY}=8$. This can be achieved by compactifying Type I or Heterotic string theory on a K3 manifold and Type IIA, IIB, M-theory and F-theory on a $CY^3$-fold. Now that we are considering non-trivial compactifications of Heterotic string theory we have to account for the interplay between the non-trivial geometry and gauge field fluxes. 

In Heterotic string theory there is a 2-form tensor field $B$ whose field strength is given by $H=dB$. The bosonic part of the low energy effective action of the Heterotic string (Heterotic supergravity) is given by
\begin{align}\begin{split}
S=&\frac{1}{2\kappa_{10}^2}\int d^{10}x \sqrt{-g}e^{-2\Phi}\left(R+4(\nabla \Phi)^2-\half |\tilde{H}_3|^2\right)\\
&-\frac{1}{2 g_{10}^2}\int d^{10}x\sqrt{-g}e^{-2\Phi}\text{ Tr}|F_2|^2~,
\end{split}\end{align}
where 
\begin{align}\begin{split}
&\tilde{H}_3=dB-\frac{1}{4}(\Omega_{YM}-\Omega_{GR})~,\\
\Omega_{YM}=\text{Tr }A\wedge dA-\frac{2i}{3}&A\wedge A\wedge A\quad,\qquad \Omega_{GR}=\text{Tr }\omega \wedge \omega+\frac{2}{3}\omega \wedge \omega \wedge \omega~,
\end{split}\end{align}
where $A$ is the 1-form gauge field and $\omega$ is the spin connection. This leads to the equations of motion
\be
dH_3=\frac{1}{2}\left(\text{Tr }R\wedge R-\text{Tr }F\wedge F\right)\quad,\qquad R=d\omega+\omega \wedge \omega~.
\ee
%

Compactifying F-theory on an elliptic $CY^3$-fold gives a six-dimensional theory with $N_{SUSY}=8$. This matches the SUSY and dimensions of Heterotic string theory on $K3$. By the conjecture above, we expect these to be dual. Consider a compactification with only trivial $H$-flux, that is the case without five-branes (sources of the B-field). This imposes the condition
\be
Tr(R\wedge R)-Tr(F\wedge F)=c_2(TM)-c_2(F)=0~,
\ee
on the gauge bundle, where $c_2(F)$ represents the second Chern class. This means that if we compactify on a non-trivial manifold, we must have instantons living in the gauge bundle.

We will now consider the extended example of Heterotic string theory on $K3$. Before we derive the duality with F-theory, we will demonstrate some physical features of this compactification. This manifold has the feature that $c_2(TK3)=24$. This means that if there are no five-branes, there must be 24 instantons living in the gauge bundle. Since the structure group of the gauge bundle is $E8\times E8$, we can have these 24 instantons divided between the two factors $E8_1\times E8_2$ as $k_1,k_2$ where $k_1+k_2=24$ \cite{Kachru:1995wm}.

Now consider the case where $[dH]\neq 0$ - that is the case of compactification with five-branes. This case is actually more clear from the Ho\u{r}ava-Witten construction by taking M-theory on $\IR^{1,5}\times K3\times S^1/\IZ_2$. At the ends of the $S^1/\IZ_2$ interval, there are M9-branes - each of which support an E8 gauge group. Instantons in the M9-branes are described by dissolved M5-branes. These instantons in the gauge theory can shrink to zero size in the $K3\times S^1/\IZ_2$ direction and eject from the M9-brane as M5-branes into the bulk of the $S^1/\IZ_2$ (whose worldvolume is transverse to the $K3\times S^1/\IZ_2$) and can also be absorbed by the other wall and dissolved in it.  The point is that there must be a total of 24 M5-branes for this compactification to be stable spread out between the two M9-branes and the bulk \cite{Witten:1995gx}. 

Now using the duality conjecture, we know that the compactification of Heterotic string theory on $K3$ is a 6D theory with $N_{SUSY}=8$. This can also be achieved by considering F-theory on an elliptically fibered $CY^3$ manifold. As we will see this can be derived from the previous subsection by applying the `adiabatic argument'.

 Let us start with the duality of F-theory on K3 and Heterotic string theory on $T^2$. Recall that $K3$ can be written as an elliptic fibration 
\be
\xymatrix{T^2\ar[r]&K3\ar[d]\\&\IP^1}~.
\ee
Now if we take both sides of the duality and fiber them over a 2-sphere $\IP^1$, then we get
\begin{equation}
\text{F-theory on }
\vcenter{\vbox{
\xymatrix{K3\ar[r]&CY^3\ar[d]\\&\IP^1}}}\cong \vcenter{\vbox{\xymatrix{T^2\ar[r]&CY^3\ar[d]\\&\IP^1\times \IP^1}}}\Longleftrightarrow\quad \text{Heterotic on }\vcenter{\vbox{\xymatrix{T^2\ar[r]&K3\ar[d]\\&\IP^1}
}}~.
\end{equation}

Here $CY^3$ is a Calabi-Yau 3-fold which is $K3$ fibered over  $\IP^1$.  For the case $k_1=k_2=12$ this corresponds to a $CY^3$ which is elliptically fibered over $\IP^1\times \IP^1$ \cite{Kachru:1995wm}.  For the more general splitting $k_1=12-n, k_2=12+n$ the the base $\IP^1\times \IP^1$ is replaced by a Hirzebruch surface $F_n$
\cite{Morrison:1996na, Morrison:1996pp} .  The comparison between the Higgs branches of Heterotic side and its geometric interpretation on the F-theory side can be found in
\cite{Bershadsky:1996nh}.
In this duality, the transition of an instanton from one $E8$ factor to the other has an interesting geometric interpretation on the F-theory side.  Emitting an instanton to the bulk 
corresponds to blowing up a point on $F_n$ and absorbing it on the other side is equivalent to blowing another $\IP^1$ down to a point and the net effect is converting $F_n$ to $F_{n\pm1}$ \cite{Witten:1996qz}.

\subsubsection{$N_{SUSY}=4$ and 4D theories}
\label{sec:4D}
Similarly we can continue down to theories with 4 supercharges.  The highest dimension for which this happens is ${\cal N}=1$ supersymmetric theories in four dimensions.
We can obtain these theories compactifying by Type I or Heterotic strings on Calabi-Yau 3-folds,  F-theory on Calabi-Yau 4-folds, or M-theory on G2 holonomy manifolds.  Needless to say we expect these to be dual to one another with suitable choices of parameters.  There has been a large and growing literature on this subject which is beyond the scope of the current review.  A particularly powerful description of this class of theories involves F-theory, as it is based on relatively simple geometrical data.  This description has led to many potential connections with supersymmetric particle phenomenology involving supersymmetric extensions of the standard model. See \cite{Heckman:2010bq} for a partial review.  

Given the large number of six-dimensional manifolds, we would like to know something more about what kinds of compactifications are allowed. Generic compactifications of this type can have a large amount of matter fields depending on the number of non-trivial homology cycles. As we go to large number of matter fields in any theory, we run into the problem of the theory being not asymptotically free. In general this means that we cannot decouple gravity. The question now becomes what constraints does this place on our theory, or in other words, can the non-asymptotically free theory be arbitrary? As it turns out the restrictions are quite strong. 

Since the standard model is asymptotically free\footnote{We assume that it is embedded in some grand unified theory.},  we can decouple gravity even though we do have gravity in our universe (and hence have a compact internal manifold assuming this is the correct description). In order to study just the gauge theory part of this, we can study a non-compact version of the internal manifold, $Y_{nc}$, which describes the local structure of the true compact manifold, $Y_c$. In essence we can think of this as ``zooming'' into a local part of $Y_c$ so that it appears to be so large that we can think of it as a non-compact manifold.   One of the lessons one learns in this context is that local singularities of CY geometry play a key role in encoding and restricting some aspects of phenomenology.  One particularly promising example of this is how flavor hierarchy can be geometrically encoded \cite{Heckman:2008qa}. Given the importance of local singularities we turn to a brief review of it later in this lecture.

An important difference between compact and non-compact Calabi-Yau manifolds is that compact manifolds cannot have global symmetries while non-compact ones can. This means that the global symmetries of the standard model (decoupled from gravity) can only be approximate symmetries when we take gravity into account, matching beautifully with the first swampland conjecture in Lecture \ref{sec:Lec2} \cite{Vafa:2005ui}.

Of course we need to understand the full compact geometry in order to understand the structure of the complete standard model, however studying a non-compact version should be sufficient for some low energy approximations which we can think of as only probing the local geometry of $Y_c$. It is also worth pointing out that not every local model of this type is permissible. These can lead to contradictions with observations and consistency and hence should be counted as being in the ``swampland'' of the string landscape -  the part that appears to be a consistent low energy theory which does not have a consistent UV completion with gravity. This will be the topic of Lecture \ref{sec:Lec2}.  

\subsubsection{Supersymmetry Breaking}

At the time of writing these notes, supersymmetry had not yet been realized at energy scales tested by collider experiments. However, supersymmetry is a fundamental component of string theory model building
. Therefore, in order to properly describe real physics at low energy, we must somehow figure out how to break supersymmetry. This has been a topic of intense study for a long time which is deceptively difficult. For example, consider the following two cases

\begin{enumerate}
\item Compactify a supersymmetric theory on $T^2$. We can then break supersymmetry by hand by imposing anti-periodic boundary conditions for fermions. 
This is called the Scherk-Schwarz compactification \cite{Scherk:1979zr}. One drawback of this approach is that this theory will develop a tachyon for a small enough radius of the circle compactification. This problem is unavoidable because we cannot study such a system and ignore the radial modulus, or arbitrarily restrict attention to only a subset of the radion moduli space \cite{Rohm:1983aq}.
\item 
We can break supersymmetry by considering compactifications on manifolds without a special holonomy.  For example if compactify on two complex dimensions and if we have a local orbifold singularity of the type $\IC^2/\Gamma$, where $\Gamma \not\subset SU(2)$ this breaks SUSY.  But again, studying string theory in this background we learn we have tachyons and hence the system is unstable \cite{Vafa:2001ra}.
\end{enumerate}
 This tachyonic behavior and more generally lack of stationary solutions appears to be a ubiquitous behavior when we try to break supersymmetry. It would appear that string theory is sending us a message. 

A notable proposal to break supersymmetry and reduce to our universe with positive cosmological constant is given by KKLT \cite{Kachru:2003aw}. This approach broadly consists in two steps: 1.) stabilizing the moduli while preserving supersymmetry and 2.) breaking supersymmetry by adding anti-D3-branes wrapped on highly warped cycles in the internal manifold. We will now review this construction in more detail.

Consider F-theory/IIB supergravity on a Calabi-Yau orientifold with fluxes, with the simplifying assumption of having only one K\"ahler modulus. Due to the scale invariant property of the effective Lagrangian, the complex structure moduli can be stabilized perturbatively while the K\"ahler moduli cannot. Therefore in order to stabilize all moduli one must add non-perturbative features such as Euclidean D3 instantons \cite{Witten:1996bn}.
Once all moduli are stabilized, we will add anti-D3-branes. It should be noted that those anti-D3-branes will not bring other moduli, as their worldvolume scalars are all automatically stabilized by the fluxes. These anti-D3-branes are then wrapped on a cycle at the tip of a Klebanov-Strassler throat \cite{Klebanov:2000hb}, inside the Calabi-Yau orientifold. The anti-branes will then back react on the non-compact part of spacetime, causing it to have a positive cosmological constant, much like how D3-branes lead to $AdS$ space. In order to properly understand the KKLT construction, one must have full control of the large number of moduli which is generally intractable. See \cite{Moritz:2017xto} for related issues in the KKLT scenario.

We are now at an impasse. It might be that the KKLT construction consistently breaks SUSY and reproduces a $dS$-like vacuum with positive energy, but we are just unable to analyze it yet. Since we do not have the tools to analyze the KKLT construction in a realistic case, it is impossible to make any strong claims about a KKLT-like realization of $dS$ in string theory. Indeed there are various problems which arise in this class of theories. In introducing anti-D3-branes on a compact manifold, we need to take special care of having them well separated from the D3-branes, so that they do not annihilate. Therefore, to analyze stability when introducing both D3-branes and anti-D3-branes, we really must have a complete knowledge of the moduli space, including both open and closed string moduli. This is clearly a very complex problem, as the question of moduli stabilization remains an active area of research to date.


\subsection{Singularities and Branes}
Another important facet of our modern understanding of string theory is the role of singularities.  
This is demonstrated in an example from the previous section where Type IIA on K3 is dual to Heterotic string theory on $T^4$. In this example, non-abelian gauge symmetry on the Heterotic side is dual to the K3 manifold developing singularities. It is clear from this that the singularities must play an integral part of understanding these physical theories.   This is just an example of the general principle that singularities give rise to interesting physical phenomena.   Moreover understanding local singularities can lead to a deeper understanding of quantum systems, decoupled from gravity.  This in particular has led to insights about the existence of non-trivial interacting quantum systems in up to six dimensions.
 Before we demonstrate some of the interplay between singularities and branes, it is helpful to do a quick review of toric geometry.

\subsubsection{Lightning Review of Toric Geometry}

Toric geometry is the study of toric varieties. A \emph{toric variety} $X$ is the zero set of a collection of complex polynomial equations that have an algebraic torus $T=(\IC^\ast)^r$ as a dense open subset and has a natural action action $T:X\to X$ such that the restriction to $T\subset X$ is the usual translation action \cite{Hori:2003ic}. 
\begin{figure}[t]
\centering
\includegraphics[scale=0.9,trim=3cm 17cm 6cm 5.6cm,clip]{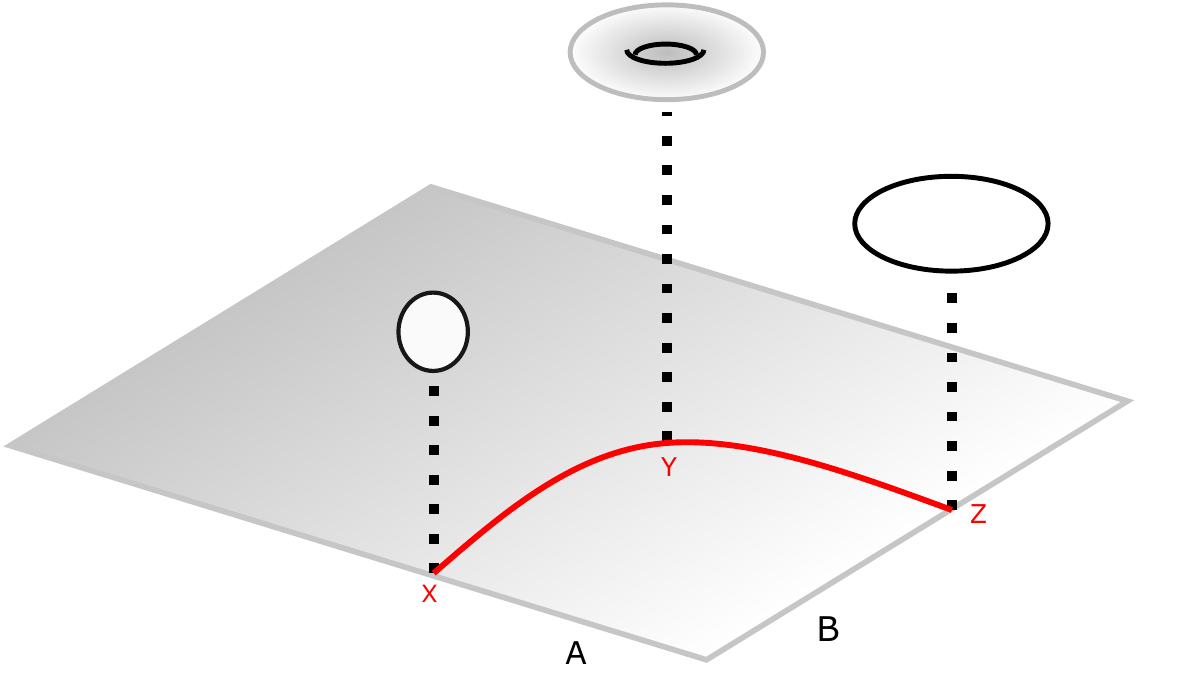}
\caption{In this figure we illustrate the idea of toric varieties. Here the two boundaries (labeled A, B) are where the A- and B-cycles of the $T^2$ torus fiber degenerates. So the fiber above points X, Y, and Z will be the $S^1$ B-cycle, full $T^2$, and $S^1$ A-cycle respectively.}
\label{fig:FiberDeg}
\end{figure}

\begin{figure}[b]
\begin{center}
\includegraphics[trim=0.5cm 25cm 11.5cm 1.5cm,clip]{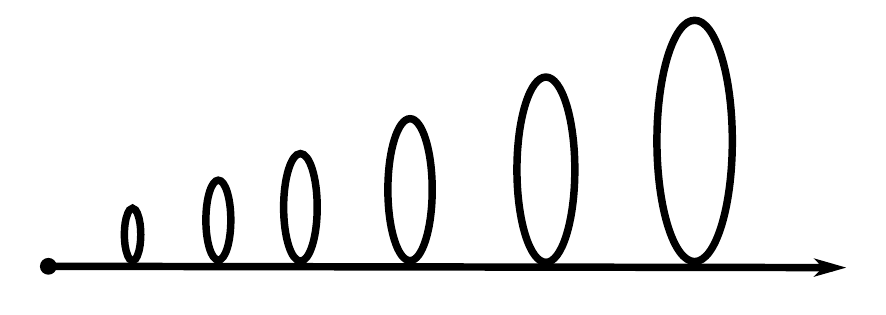}
\caption{This figure is a representation of how $\IC^1$ can be realized as a toric variety which is a circle fibered over the positive real line. }
\label{fig:C1}
\end{center}
\end{figure}

A natural way to encode the topological data of a toric variety is by realizing them as torus fibrations. Since we can realize a generic toric variety $X$ as 
\be
\xymatrix{T^r\ar[r]&X\ar[d]\\&B_r}~,
\ee
with a $T^r$ fiber, we can realize the base of the fibration as a polyhedra where different cycles of the fiber degenerate along the different boundary components $\partial B_r=\coprod_{i}\sigma_i$ where the number of degenerate cycles on $\sigma_i$ is given by codim$_{B_r}~ \sigma_i$. See Figure \ref{fig:FiberDeg}. When $\partial B_r$ forms a closed polyhedra, we can encode its data in the dual graph -- this dual graph is called the \emph{toric fan} and is related to the way in which branes arise in the Type IIB description of M-theory compactified on a toric variety \cite{Leung:1997tw}.

Consider the example of $\IC$. This can be realized as a plane where the complex coordinates $(z,\bar{z})$ can be exchanged for the real coordinates $(r,\theta)$ in the usual way. This allows us to realize $\IC$ as a a circle ($\theta$-coordinate) fibered over the positive real line, $\IR_+$, ($r$-coordinate) where the radius of the circle is given by $r$. See Figure \ref{fig:C1}. 

\begin{figure}
\centering
\includegraphics[scale=0.9,trim=2cm 15cm 4cm 5cm,clip]{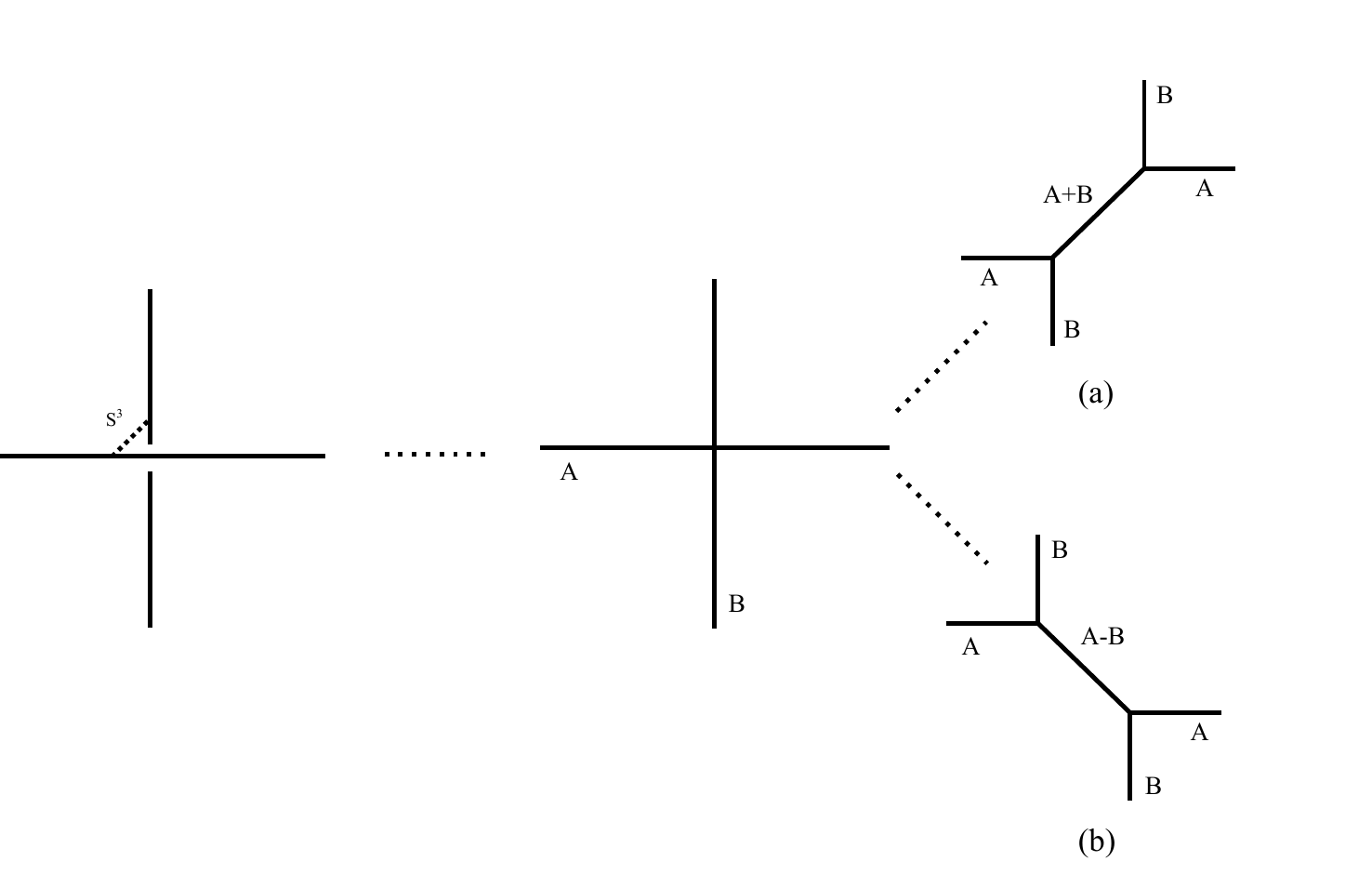}
\caption{This figure shows the two possible resolutions ( left and right) of the conifold (center). In the conifold, the vertical and horizontal line represent the locus in the base where two perpendicularly cycles (A- and B-cycles) in the $T^2$ fiber degenerate. This means that the preimage of a line running diagonally from one axis to the other in the base represents a 3-sphere since it is a 3-real dimensional manifold with a perpendicularly embedded $S^1$ degenerating at each end. This can be resolved in two different ways which corresponds to resolving with an $S^3$  (left) or by gluinga $\IC\IP^1$ at the origin (right) where the degenerate $S^1$ fiber is either the (a) A+B cycle or the (b) A-B cycle in the $T^2$ fiber.}
\label{fig:conRes}
\end{figure}

 In this way, we can similarly realize $\IC^3$ as 3-perpendicular copies, that is a $T^3$ fibered over $\IR_+^3$ which looks like an octant in $\IR^3$ where the radius of the different cycles of the torus are determined by their distance from the 3 axes. 

An important feature of the description of toric varieties as torus fibrations of a base polytope
 is that it has a clear geometric interpretation of blow ups. The geometric operation of \emph{blowing up} is a method of resolving a singularity by replacing a singularity with a smooth manifold. In some sense, the most fundamental blow up replaces a singularity with a copy of $\IC\IP^1$. More generally singularities can be replaced by a collection of $\IC\IP^1$'s which can be glued together in interesting ways to form more complicated resolutions, or by other manifolds.   As a toric variety, $\IC\IP^1$ can be realized as a $S^1$ fibered over a finite line segment where the fiber degenerates at the two ends. Using this, the fundamental resolution of singularity simply replaces the singular point where cycles degenerate on the base with a line segment where the fiber is given by a linear combination of the degenerating cycles at the unresolved singularity. See Figure \ref{fig:conRes} for an example. See \cite{Leung:1997tw,Hori:2003ic,Iqbal:2003Ds} for more details. 

\subsubsection{The McKay Correspondence and Theories of Class $\mathcal{S}$}
\label{sec:McKay}
Let us study the case of K3 singularities in more detail. The allowed singularities of K3-manifolds are locally of the form $\IC^2/\Gamma$ where $\Gamma\subset SU(2)$ is a finite group. These generally have an ADE classification\footnote{That is they have a classification which is identical to that of A-,D-, and E-type Lie algebras.}. These are given by
\begin{itemize}
\item A-type: $\Gamma$ is a binary cyclic group, $\Gamma=\IZ_n$
\item D-type: $\Gamma$ is a binary dihedral group, $\Gamma=BD_{2n}$
\item E6-type: $\Gamma$ is the binary tetrahedral group $\Gamma=2T$
\item E7-type: $\Gamma$ is the binary octahedral group $\Gamma=O$
\item E8-type: $\Gamma$ s the binary icosahedral group $\Gamma=2I$
\end{itemize}
To illustrate these we will consider the $A_{n-1}$ type singularity in detail. Consider the group $\Gamma=\IZ_n$ which is defined by the generator
\be
\Gamma=\left\langle \left(\begin{array}{cc}\alpha&0\\0&\alpha^{-1}\end{array}\right)\right\rangle\quad,\qquad \alpha^n=1~,
\ee
acting on $\IC^2$. Roughly speaking, Type IIA string theory on $\IR^{1,5}\times\IC^2/\IZ_n$ is T-dual to Type IIB with $n$ coincident D5-branes giving rise to a $SU(n)$ gauge theory \cite{Ooguri:1995wj,Katz:1996ht}\footnote{There is a slight subtlety associated with this. We technically need to have a singular Taub-NUT space, $TN_n$, (the singular limit of $n$ NUT centers) which locally has the same singularity structure as $\IC^2/\IZ_n$. The difference between $TN_n$ and $\IC^2/\IZ_n$ is that in $TN_n$, the circle fiber asymptotically 
approaches a finite so that T-duality is well defined whereas the $S^1$ fiber of $\IC^2/\IZ_n$ diverges as the distance from the singularity. Therefore, T-duality is not technically defined in $\IC^2/\IZ_n$ as one cannot shrink the asymptotic circle to apply T-duality \cite{Gregory:1997te,Witten:2009xu}.  There is however, a version of mirror symmetry which work for this case as well \cite{Hori:2000kt}.}. This $SU(n)$ gauge theory can be seen in the Type IIA side by blowing up the singularity: completely resolving the singularity results in a collection of $\IP^1$s that intersect as in the $A_{n-1}$ Dynkin diagram\footnote{Technically, there are $2^{n-1}$ different spheres in the fully resolved $\IC^2/\IZ_n$, $\widehat{\IC^2/\IZ_n}$, and we can choose a basis for the homology group $H_2(\widehat{C^2/\IZ_n})$ such that their intersection matrix is given by the $A_n$ Cartan matrix.}. The gauge theory then comes from D-branes wrapping these non-trivial 2-cycles which upon going to the singular limit, all become effectively coincident, giving rise to a non-gauge symmetry. 

We can also get the $D_n$ type theories in a similar way. This corresponds to having orientifolds in the Type IIB picture and in the geometric picture by replacing $\IZ_n\to BD_{2n}\cong\IZ_n\rtimes \IZ_2$ where the $\IZ_2$ action acts as an orbifold. Understanding the interpretation of E-type singularities as branes in Type IIB requires the use of F-theory since the branes have do not have a perturbative description.  It is known in the M-theory setup or the Type IIA picture that E-type singularities lead to E-type gauge theories. 

One can then fiber these geometries as in the adiabatic construction and obtain systems with lower supersymmetry in lower dimensions.  In particular fibering these geometries over $\IP^1$ leads to ${\cal N}=2$ supersymmetric systems in 4 dimensions.  Using this picture and applying local mirror symmetry, the Seiberg-Witten curve of these theories can be identified in string theory \cite{Klemm:1996bj} as in Appendix \ref{sec:App4D}. As we will now explain, this curve can also be realized in Type IIA as the geometry of intersecting D4/NS5-branes, which lift to M5 branes in the M-theory lift of Type IIA \cite{Witten:1997sc,Gaiotto:2009we}.

If instead of Type IIA, we consider the compactification of Type IIB string theory on a K3-manifold with $\IC^2/\Gamma_\fg$ singularities, this leads to a six dimensional theory with $\CN=(2,0)$ with type $\fg$ conformal systems \cite{Witten:1995zh,Claus:1997cq,GMooreFelixKlein,Nahm:1977tg}. Compactifying these theories on a Riemann surface $C$ (with a topological twist) 
produces the four dimensional $\CN=2$ theories of class $\CS$ \cite{Gaiotto:2009hg,GMooreFelixKlein,Witten:1997sc,Gaiotto:2009we}. These theories can be constructed for type $\fg=A_{N-1}$\footnote{Note that this construction can be generalized to $\fg=D_n$ type theories by wrapping an $O5$-plane on $\Sigma\times \IR^4$.} by compactifying $N$ M5-branes on a Riemann surface $\Sigma\times \IR^4$ with the same topological twist, where $\Sigma\to C$ is a multisheeted cover of $C$. This derives the data of the Seiberg-Witten curve and Seiberg-Witten 1-form, where the curve the Prym variety\footnote{The Prym variety is the kernel of the induced map $\pi^\prime:J(\Sigma)\to J(C)$ from $\pi:\Sigma\to C$. Here $J(X)$ is the Jacobian variety of $X$ which is defined as the quotient of all global holomorphic differentials on $X$, $H^0(\Omega_X^1)$, quotiented by the space of non-trivial closed 1-cycles on $X$, $H_1(C)$. By the universal covering construction, this space has the same $1^{st}$ homology group $H_1(X)=H_1(J(X))$.} associated to the map $\Sigma\to C$ \cite{Witten:1997sc,Gaiotto:2009we,Gaiotto:2009hg}. In addition, this gives a clear geometric way to study the BPS spectrum, line operators, surface operators, and expectation values of supersymmetric operators in the associated four dimensional theory \cite{Shapere:1999xr,Alday:2009aq,Gaiotto:2012rg,Gaiotto:2011tf,Gaiotto:2010be,Gaiotto:2012Db,Alim:2011kw,Alim:2011ae,Gabella:2017hpz}. See \cite{Tachikawa:2013kta,Teschner:2014oja,Gaiotto:2014bja,GMooreFelixKlein} for a more general review.  

\subsubsection{Branes in F-Theory}

The relation between singular geometry and branes is also evident in F-theory. If we take F-theory on $K^{ell}$ where
\be
\xymatrix{T^2\ar[r]&K^{ell}\ar[d]\\&B}
\ee
this is dual to Type IIB on $B$ with $(p,q)$ 7-branes where the $(p,q)$ cycles of the $T^2$-fiber degenerate. A choice of $(p,q)$ cycles corresponds to a choice of S-duality frame for the Type IIB theory. These branes are non-perturbative because they are co-dimension 2 and hence the magnetic dual of the axio-dilaton has log-type asymptotic behavior. In general these branes are mutually non-local due to their $(p,q)$ axion, dilaton charges and can give rise to non-abelian exceptional groups as studied in  \cite{Gaberdiel:1997ud}.  One may worry about having these charged objects in a compact space because branes source flux and as we know from general theorems of general relativity we cannot have a net charge in a compact space as the flux has nowhere to go. However, this F-theory setup avoids this problem because the 7-branes source non-abelian flux which can actually cancel with themselves \cite{Vafa:1996xn} and thus lead to realization of non-trivial stable charged objects in a compact space.

\subsubsection{5-Brane Webs}
\begin{figure}
\begin{center}
\includegraphics[scale=0.5]{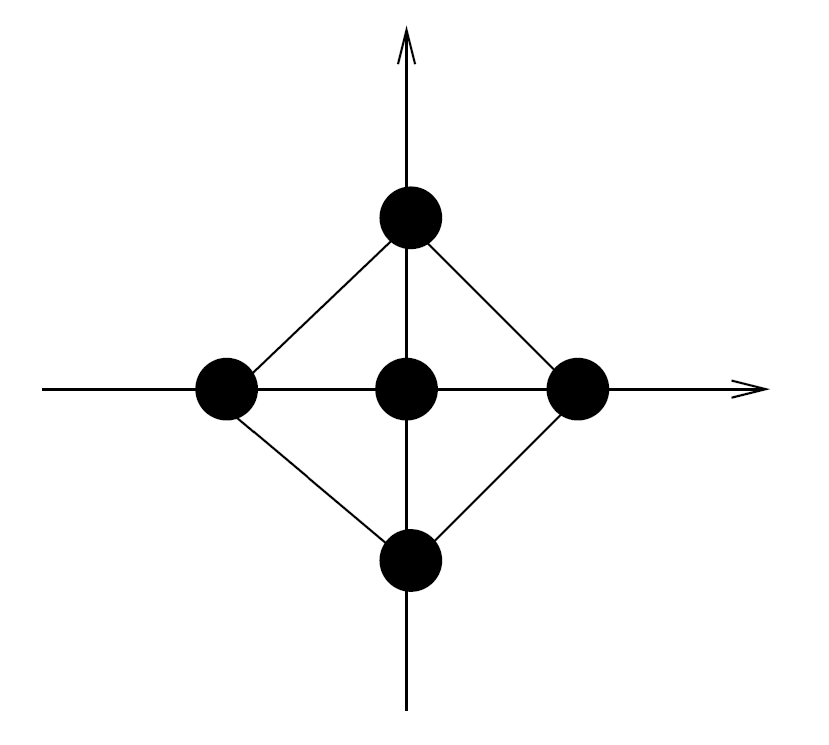}
\caption{This is the toric diagram for the $CY^3$-fold given by the affine cone over $\IP^1\times \IP^1$: $\CO(-2,-2)\to \IP^1\times \IP^1$. This diagram encodes the data of the singular structure of the $T^2$ fiber via its dual graph. }
\label{fig:toric}
\end{center}
\end{figure}
Another  example of the correspondence between branes and geometry is the construction of brane webs \cite{Aharony:1997bh}. Consider M-theory on $\IR^{1,6}\times K^{ell}$ where $K$ is an elliptically fibered $$CY^3$$. If we compactify and T-dualize on a pair of orthogonal cycles on the $T^2$ fiber, we get a Type IIB theory with $(p,q)$ 5-branes which have $p$ R-charge ($\tilde{C}_2$) and $q$ NS-charge\footnote{Here $\tilde{B}$ represents the magnetic dual of the form field $B$ and similarly for $\tilde{C}_2$ and $C_2$.} ($\tilde{B}$) where we compactify along the $(1,0)$-cycle and T-dualize along the $(0,1)$-cycle. This generalizes the relation of $n$ coincident D6-branes in Type IIA to Taub-NUT geometry (with $\IC^2/\IZ_n$ singularity) by lifting to M-theory and the relation of Taub-NUT to NS5-branes by T-duality \cite{Leung:1997tw}. 

Let us consider an explicit example. Consider M-theory on the non-compact space 
\be
\xymatrix{T^2\ar[r]&\CO(-2,-2)\ar[d]\\&\IP^1\times \IP^1}
\ee
On the base $\IP^1\times \IP^1$, the singular structure of the the $T^2$ fiber can be represented by the dual of the toric diagram in Figure \ref{fig:toric}.

This means that after dualizing to Type IIB, we have a system of $(p,q)$ 5-branes on $\IR^{1,4}\times \IR^2_{web}\times \IR^3$ where the 5-branes wrap the $\IR^{1,5}$ direction and form the generically co-dimension 1 web in the $\IR^2_{web}$ direction as shown below:

\begin{figure}[h]
\centering
\begin{tikzpicture}
\draw [-, very thick, black] (0,-1)--(0,1);
\draw [-, very thick, black] (2,-1)--(2,1);
\draw [-, very thick, black] (0,1)--(2,1);
\draw [-, very thick, black] (0,-1)--(2,-1);
\draw [-, very thick, black] (2,1)--(3,2);
\draw [-, very thick, black] (0,1)--(-1,2);
\draw [-, very thick, black] (0,-1)--(-1,-2);
\draw [-, very thick, black] (2,-1)--(3,-2);
\node (pqlab1) at (1,1.25) {$(1,0)$};
\node (pqlab1) at (1,-1.25) {$(1,0)$};
\node (pqlab1) at (2.5,0) {$(0,1)$};
\node (pqlab1) at (-0.5,0) {$(0,1)$};
\node (pqlab1) at (3.5,2) {$(1,1)$};
\node (pqlab1) at (-1.5,2) {$(1,1)$};
\node (pqlab1) at (3.5,-2) {$(1,1)$};
\node (pqlab1) at (-1.5,-2) {$(1,1)$};
\end{tikzpicture}
\end{figure}

In essence, this brane web depicts two D5-branes suspended between two NS5-branes. This  describes the 5D $\CN=2$ $SU(2)$ SYM where the displacement between the D5-branes (vertical displacement $x^6$) gives the Higgs vev and the displacement between NS5-branes (horizontal displacement $x^5$) determines the gauge coupling as $
\Delta x^5=\frac{4\pi}{g_{YM}^2}$.

One can also combine branes and singularities, by bringing branes to probe singularities and obtain new and interesting quantum systems.  A brief review of some examples is given in Appendix \ref{app:A}.

\section{Lecture 2: The Swampland}\label{sec:Lec2}

In the previous lecture we briefly reviewed the string theory construction of many low dimensional effective field theories. As we saw, there is a very large number of choices to make when using string theory for model building, coming from the choice of the compactification manifold, background fluxes, branes, and etc. Therefore a very relevant question is how to identify which particular string theory solution, among the enormous set of possibilities, describes our universe. 

It has recently been estimated that the number of possible consistent flux compactifications of F-theory to $4D$ is at least $10^{272000}$ \cite{Taylor:2015xtz}. Although it is unknown if all of these compactifications are distinct or may be dual descriptions of the same theory, this large number of string vacua suggests that perhaps the direct study of all string vacua is futile. This remarkably large space of inequivalent string backgrounds is called the \emph{string landscape}. 

To complicate matters even more, even if we were to be able to enumerate all of the distinct, consistent string backgrounds, there is no known top-down mechanism to prefer one particular choice over another. For example, what forces four dimensions to be extended and six to be compact? Why did nature choose the specific string background describing our universe from the vast number which can be constructed in the theory? While there have been some suggested ideas \cite{Brandenberger:1988aj,Kofman:2004yc}, there is no compelling solution yet.

Due to this huge number of possible choices involved in constructing string vacua, there has been a distinct philosophical shift in the community over the past decade. The attitude towards identifying ``the correct'' string vacuum has shifted from using a top-down approach to a bottom-up one. Instead of starting with fully-fledged string theory and studying the compactifications down to 4D, many have started studying effective four dimensional quantum field theories with nice phenomenological features (such as supersymmetric extensions of the standard model) and then trying to couple them to gravity. The common lore is that because the string landscape is so large, it is likely that any consistent looking lower dimensional effective field theory (EFT) coupled to gravity can arise in some way from a string theory compactification.   Indeed this idea would make string theory pretty much irrelevant for phenomenological questions.

In this lecture, we aim to argue that this way of thinking is incorrect -- that not all consistent looking EFTs can be coupled consistently to gravity with a UV completion. Since string theory is the only known UV complete theory of gravity\footnote{One could make the argument that there are other theories of quantum gravity different from string theory, such as Vasiliev higher spin theories\cite{Vasiliev:1995Dn} or even loop quantum gravity\cite{Smolin:2004sx}. However, it has been conjectured that all these other theories can be attained as a special limit of string theory\cite{Dijkgraaf:2004te}.}, we will necessarily demand that these theories arise from some string theory compactifications \cite{Vafa:2005ui}. 
The rest we believe are ultimately inconsistent.   In analogy with the string landscape, we will call the set of all EFT which do not admit a string theory UV completion as the \emph{swampland}.

It is therefore crucial to understand if a given EFT coupled to gravity lies in the string landscape or the swampland. In order to do so, we would like to identify a complete set of \emph{swampland criteria} which will identify if an EFT admits a string theory UV completion or not.  
Thus far, we have a conjectured, minimal criteria that allows us to exclude a theory from the string landscape. 
In this lecture we plan to briefly discuss ten swampland criteria. We will be unable to provide proofs, but rather will provide physical reasoning based on realization in string theory and general facts about qunatum gravity to motivate each of the conjectured criteria.  The criteria we present here are based on \cite{Vafa:2005ui,Ooguri:2006in,ArkaniHamed:2006Dz,Ooguri:2016pdq}.

\subsection{No Continuous Global Symmetries}

\noindent\emph{An effective field theory coupled to gravity cannot have (continuous or discrete) global symmetries.}\\

The motivation for this criterion relies on black hole physics. Suppose we have an EFT coupled to gravity which has a global symmetry $G$. In the spectrum of the EFT we will have states charged under such global symmetry. Now send a state charged under a global symmetry inside a black hole. The information of this global symmetry is lost by the no-hair theorem. Thus, when the black hole evaporates via Hawking radiation \cite{Hawking:1974sw} it will do so by emitting particles which carry equal number of positive and negative charges under $G$ since there is no imprint of global charges on a black hole.  This process would then violate charge conservation in $G$
as we started with non-zero charge and all the charge has disappeared after the evaporation of the black hole and no net charge has come out.

The only way to avoid this seeming contradiction is by forbidding any theory of quantum gravity from having global symmetries. Remarkably, it appears that string theory already knows about this criterion, as in all examples we know, all global symmetries are actually gauged. This is true because usually global symmetries in EFTs obtained by string compactification arise from symmetries of the extra dimensions, but such symmetries are gauged since diffeomorphisms of the compactification manifold are part of the gauge symmetry of gravity.
 
\subsection{All Charges Must Appear}

\noindent\emph{A consistent effective field theory with gauge group $U(1)$ coupled to gravity must have states with arbitrary charge $\IZ$.}\\

 A nice discussions of this condition for swampland is given in \cite{Banks:2010zn}. 
Suppose we have a $U(1)$ gauge symmetry in the EFT. The Hilbert space of the EFT will be split into different sectors, one for each value of the $U(1)$ charge of the states in that sector. Now, if the theory is not coupled to gravity, it is possible that the spectrum contains only states of some specific subset of charges, or even maybe no charged states at all. However, this cannot happen once the theory is coupled to gravity. 

   Consider a $U(1)$ theory coupled to gravity. We will have charged black hole solutions of Einstein's equation for any integral charge $Q$. By Hawking's formula, we know the black hole entropy is given by
\begin{equation}
S=\dfrac{A}{4G}~.
\end{equation} Such an entropy must have a statistical mechanics interpretation as a sum over the black hole microstates. As the black hole is charged, such microstates must be charged. Therefore all charged states should exist in the spectrum\footnote{This argument is  valid as long as charge is large enough so we can truly interpret the object as a black hole.  In other words, as long as the area of the horizon is much bigger than the Planck scale.}.

Notice that this criterion rules out many simple quantum field theories, as for example pure Maxwell theory coupled to Einstein gravity.  Such a theory must also have (perhaps massive) charged particles.  Note that at this level we did not say anything about the masses of this infinitely many charged states which should belong to the spectrum.

\subsection{Finite Number of Massless Fields}

\noindent \emph{A $d$-dimensional EFT coupled to Einstein gravity must have a finite number of massless fields. 
Moreover, the number of massless fields is bounded from above by a certain number $N_{max}(d)$ which depends only on the number of spacetime dimensions $d$.\\}

The motivation for this criterion is based on supersymmetric examples. Massless scalar fields in a lower dimensional EFT are generated in string theory by compactification. For example if we compactify on a K\"ahler manifold in order to preserve some supersymmetry in the lower dimensional EFT, the number of scalars will generically be proportional to specific Hodge numbers of the compactification manifold. For the case of compact CY manifolds, there seems to be an upper-bound on the possible hodge numbers, even though there is no proof of this.

Remarkably, string theory seems to ``be naturally aware" of this fact, and seems to have ways for preventing us to get consistent lower dimensional EFT with arbitrarily large number of light scalar fields. An easy example in which we can see this at work is the following. Consider Type IIA on $\mathbb{C}^2/\mathbb{Z}_N$. We saw in the first lecture that in this way we can realize an $SU(N)$ gauge group. There are therefore $N^2-1$ massless gluons in the spectrum. At this level there is no bound on $N$, which we can take as large as we want, therefore having an arbitrary high number of gluons. This is in no contradiction with the conjecture stated above, since $\mathbb{C}^2/\mathbb{Z}_N$ is non-compact and therefore gravity is decoupled in the EFT. 

However, in order to couple this $SU(N)$ gauge theory to dynamical Einstein gravity, we need to embed $\mathbb{C}^2/\mathbb{Z}_N$ into a compact manifold, i.e., K3.  Quite remarkably, what happens in this case is that in order for the compactification to be consistent, it must have $N\leq 20zx$, therefore putting an upper bound on the number of gluons \cite{Aspinwall:1995zi}.  See Section \ref{sec:McKay} for more details. 

Note that since we are assuming only Einstein gravity, we exclude more exotic gravity theories with infinite number of massless fields such as Vasiliev theory  \cite{Fradkin:1987ks}. 

\subsection{No Free Parameters}

\noindent\emph{A consistent EFT coupled to gravity must have no free parameters. Every parameter entering in the Lagrangian should be viewed as the vacuum expectation value of a field.}\\

Notice that this criterion puts a lower bound on the number of possible scalar fields, therefore being complementary to criterion number 3.
Again the motivation is that this appears to be true in string theory.  For example M-theory in 11 dimensions has no free parameter.  10 dimensional superstrings would naively appear to have a free parameter given by the choice of the coupling constant, but upon closer inspection one find out that the coupling constant is the expectation value of a scalar field called the dilaton.  When we compactify to go to lower dimensions we end up with effective theories whose parameters get related to the internal geometry of the compactification which again can be viewed as part of the dynamical degrees of freedom of the theory.


\subsection{The Moduli Space is Non-Compact}

\noindent\emph{The moduli space $\mathcal{M}$ of vacua (if non-trivial) 
is non-compact. In more detail, fix a point $p_0 \in \mathcal{M}$. Then $\forall\ T>0\, , \exists~ p\in \mathcal{M}$ such that \begin{equation}
d(p_0,p)>T~.
\end{equation}
where $d(p_0,p)$ is the distance between $p_0$ and $p$, computed by using the moduli space metric as the length of the geodesic passing through $p$ and $p_0$.}\\

To elucidate this criterion, we first need to discuss what we mean by the moduli space metric.
Consider an EFT coupled to gravity, with $N$ massless scalar fields $\Phi_i$, $i=1,...,N$ with no potential. Such scalar fields arise generically in string compactification, and their vacuum expectation values $\langle\Phi_i\rangle$ is related to geometrical quantities in the compactification manifolds such as for example the volumes of some cycles or their shapes. We will call the algebraic variety parametrized by the various $\langle\Phi_i\rangle$ the \emph{moduli space} $\mathcal{M}$. 
In the EFT, the kinetic term for those scalar fields typically takes the form
\begin{equation}
\mathcal{L}_{eff}= g_{ij}(\Phi)\partial_{\mu}\Phi^i\partial_{\mu}\Phi^j+...~, \label{Leff}
\end{equation} 
where $g_{ij}$ is the metric on $\CM$. 
We can use this metric to compute distances in the moduli space, and ask if $\mathcal{M}$ is compact or not. As it turns out in all known examples from compactifying string theory, the moduli space is non-compact \cite{Ooguri:2006in}.

As an easy example in which this conjecture is realized, we can consider here the case of the moduli space of IIB supergravity. In this case, there is only the axiodilaton modulus 
$\tau$, which is the combination of the string coupling constant and a RR 0-form.
The moduli space which will be the fundamental domain of $SL(2,\mathbb{Z})$. Now, fix a point $\tau_0$ in the moduli space, and consider the length of the geodesic from $\tau_0$ to $\tau$. When we take the limit $\tau \to i\infty$ while keeping $\tau_0$ finite, the geodesic length will be approximately
\begin{equation}
T\sim\log\left(\mbox{Im}\tau/\mbox{Im}\tau_0\right)~,
\end{equation}
and we clearly see that this distance is logarithmically divergent.

\subsection{New Physics from the Boundaries of Moduli Space}

\emph{Fix a point $p_0\in \CM$. In the limit of infinite distance from $p_0$, that is as $d(p_0,p)=T\to \infty$, there will be a tower of states in the EFT whose mass decreases exponentially with $T$,
\begin{equation}
m\sim e^{-\alpha T}~.
\end{equation}}

In the previous criterion we saw that for any choice of a starting point $p_0\in \mathcal{M}$ and any real number $T>0$, we will always be able to find a (in general not unique) point $p\in \mathcal{M}$ such that the distance between $p$ and $p_0$ is larger than $T$. So in general $\mathcal{M}$ will have some non-compact directions. We want to ask now what happens when we go extremely far away in moduli space along one of those directions, or equivalently we take $T$ to be extremely large \cite{Ooguri:2006in}.

Heuristically, we can understand this by considering the point compactification of moduli space, $\hat\CM$, so that $\hat\CM$ is a finite manifold where the infinities of $\CM$ correspond to singular points of $\bar\CM$. Now going to infinity corresponds to going to a singularity where generically extra massless degrees of freedom appear. 

\begin{figure}[h!]
\centering
\includegraphics[scale=0.3]{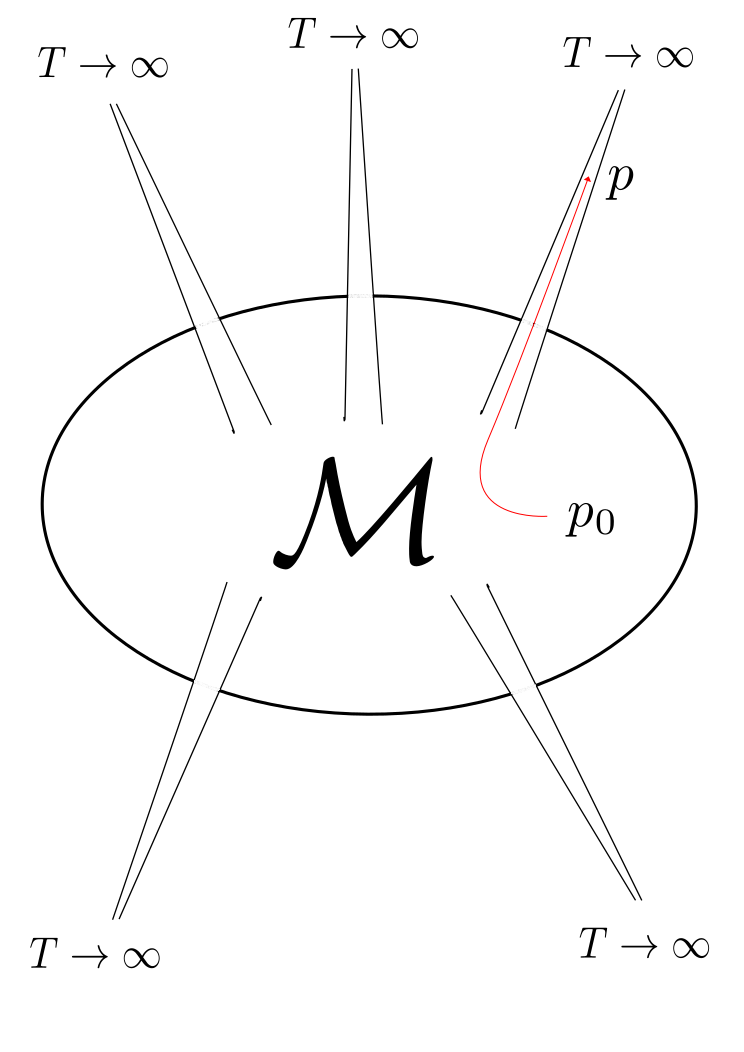}
\caption{A schematic picture of the moduli space.}
\label{Modulifig}
\end{figure}

We can illustrate this criterion in a very easy example. Consider the compactification of a EFT on a circle $\mathcal{S}^1_R$. This theory has a modulus, which is the radius $R$ of the circle. The Lagrangian for this scalar field, in one lower dimension, will be given by
\begin{equation}
\mathcal{L}_{eff}=\int \left(\dfrac{dR}{R}\right)^2+...
\end{equation}
Now, let us see how the criterion 5 and 6 apply to this case. In this example, the moduli space is just $1$-dimensional, and we can see immediately that there are infinite distances. For example, fix a radius $R_0$ and pick a $T>0$. The distance from $R_0$ to some other point $\tilde{R}$ will then be given by
\begin{equation}
\int_{R_0}^{\tilde{R}} \dfrac{dR}{R} = \log(\tilde{R})-\log(R_0)~,
\end{equation}
and we see that we can always find a suitable $\tilde{R}$ to make this distance as big as we want. So the criterion number 5 is satisfied. Let us consider the limit of very large radius, to see criterion 6 at work. As the radius $\tilde{R}$ grows to infinity, we will have Kaluza-Klein modes, with mass given by 
\begin{equation}
m\sim \dfrac{1}{\tilde{R}}~.
\end{equation}
On the other end, $T\sim \log(\tilde{R})$ and therefore we see that we have some fields with mass
\begin{equation}
m\sim e^{-T}~,
\end{equation} which get exponentially light when we go to infinity in $\mathcal{M}$.\\

\noindent{\it This conjecture also implies the remarkable fact that a consistent theory of quantum gravity must have extended objects in its spectrum.} \\

Those extended objects can be for example membranes, strings, etc. Therefore, by this criterion one can argue that quantum gravity cannot be a theory of just particles. We will now show this in the same easy example we used.
Pick now the same reference point $R_0$ but instead of going to larger radius go to smaller and smaller radius. We also find that this is another infinite distance in moduli space, as
\begin{equation}
\lim_{\tilde{R}\to 0}\int_{\tilde{R}}^{R_0} \dfrac{dR}{R}~,
\end{equation}
diverges. Therefore, due to criterion 6, we also expect to have in this case some states with mass getting exponentially low. However, such states cannot be particle states because all the particle states will be given by KK modes, and those KK modes will be instead very massive in the small radius limit.   The only way we can get light objects in the small radius limit, is to have some extended objects which can wrap around the circle which thus become lighter as we go to the small radius limit.    So if our theory does not have extended objects, we do not have any light states at all in this limit and we therefore violate the criterion number 7.  This is for example what happens in M-theory when we compactify on the circle:  The M2 branes wrapping the circle become light and give rise to the light string states in 10 dimensions.
We thus see that this conjecture implies the existence of extended objects in a consistent theory of quantum gravity.

\subsection{The Moduli Space is Simply Connected}

\noindent\emph{The first fundamental group of the closure of the moduli space is trivial, and therefore the closure of moduli space is simply connected: \begin{equation}
\pi_1(\overline{\mathcal{M}})=0~.
\end{equation}}

In all known examples in string theory, the moduli space is obtained by quotienting a contractible Teichm\"uller space $\CT$ by a group action $\Gamma$ \cite{Ooguri:2006in}
\be
\CM=\CT/\Gamma~.
\ee
In every known case, $\Gamma$ is generated by group elements which act with fixed point \cite{Ooguri:2006in}. These fixed points in $\CM$ have extended gauge symmetries given by the stabilizer in $\Gamma$. If we take $\Gamma=\langle g_i\rangle$  (group generated by $g_i$) where each $g_i$ has a fixed point in $\CM$, since $\CT$ has no non-trivial loop, the only way to get one is by the action of $\Gamma$. But since each element of $\Gamma$ can be decomposed to elements with fixed point, it implies that each loop can be contracted. This is because under the usual identification $\pi_1(\tilde\CM/\Gamma)=\Gamma$, we can identify each loop $\gamma\subset \CM$ with an element $h\in \Gamma$. Generally $h=\prod_i g_i$ where each $g_i$ has fixed points. This means that we can decompose $\gamma$ as a product of paths. Each of these path components are contractible since they can be unwinded at the fixed point. Therefore we have that $\gamma$ must also be contractible and thus  $\overline\CM$ will be a simply connected space.

\subsection{The Weak Gravity Conjecture}

\emph{In a consistent EFT coupled to gravity, gravity must always be the weakest force.}\\

This conjecture applies to charged particle states as well as charge $p$-branes.
This powerful conjecture was originally formulated in \cite{ArkaniHamed:2006Dz} and recently recieved much interest as it is able to put constraints or completely rule out different large-field inflation models. This is the so called \emph{weak gravity conjecture} (WGC). There are many inequivalent and more precise versions of this conjecture, see for example \cite{Heidenreich:2015nta,Heidenreich:2016aqi,Harlow:2015lma,Cheung:2014vva,Montero:2016tif}. Here we will present a particular version, just to give the reader the main idea.
Suppose we have a $4D$ $U(1)$ gauge theory. We already know from conjecture 2 that we need to have charged states in the spectrum of the theory. Consider then the lightest charge state, and suppose it has positive charge $q$. Consider now two of these objects together, placed at distance $r$.

There will be a repulsive electric force $F_e\sim \dfrac{q^2}{r^2}$. There will also be an attractive gravitational force $F_g\sim \dfrac{m^2}{M_p^2r^2}$. The claim is that $F_g\leq F_e$. For this to hold, it must happen that the lightest state in the spectrum satisfies \begin{equation}
\left(\dfrac{m}{M_p}\right)\leq q~.
\label{WGC}
\end{equation}
We can motivate this conjecture as follows:

\begin{enumerate}
\item It is true in our universe. The electric repulsion among two electrons, for example, is much stronger than the gravitation attraction among them. 

\item Another motivation is \emph{a posteriori}: it is always upheld in string theory constructions.
For example, we could try to violate this conjecture by making an internal manifold smaller and smaller, as we know that for example KK masses are proportional to the inverse of some geometrical size of the cycles of the compactification manifold. In this way one can get close to violating the conjecture, but then the size of the extra dimension is so small that the extra degrees of freedom become light and this description breaks down. In some way, it appears that string theory \emph{knows} about the WGC, and prevents us from violating it.

\item Another motivation for the WGC is the fact that all non-BPS black holes should be able to decay. So let us consider the example at an extremal black hole, with mass $M$ and positive charge $Q$. The extremality condition implies $M=Q$.
For this black hole to decay via Hawking radiation, it has to emit particles. But suppose now that for all the states in the spectrum we have $m>q$, then when the black hole radiates a particle it will inevitably have $M'<Q'$ after. Therefore it would violate the extremality bound, developing a naked singularity and thus also violating the cosmic censorship conjecture.
The only way out is to assume that the spectrum contains at least one particle which satisfies the bound (\ref{WGC}).

\end{enumerate}

In order to illustrate this criterion at work, we consider now an  illustrative string theory example. Take Heterotic strings on a $d$-dimensional torus. We have an equation relating the allowed masses of string excitations with momentum and winding numbers
\begin{equation}
\dfrac{1}{2}m^2=\dfrac{1}{2}P_L^2+N_L-1=\dfrac{1}{2}P_R^2+N_R~.
\end{equation}

As we will now show, the (-1) in the left moving sector is related to the inequality in the WGC. Consider first supersymmetric BPS states. $N_R=0$ and then 
\begin{equation}
m^2=P_R^2~,
\end{equation} which is the analog of $M=Q$ in the weak gravity conjecture. 
Consider now the non-supersymmetric states, in which $N_L=0$. We now have
\begin{equation}
m^2=P_L^2-1~.
\end{equation}
Again, the charge is given by $P_L^2$ and we have that this is the analog of the strict inequality in the weak gravity conjecture.

A natural question to ask now is for which states the WGC bound is saturated. The answer is given by a more sharpened version of the WGC, which is the following \cite{Ooguri:2016pdq}:

\emph{The equality sign in the Weak Gravity conjecture holds if and only if
\begin{enumerate}
\item The underlying theory is supersymmetric
\item The states saturating the WGC bound are BPS states.
\end{enumerate}}

A very nice application of the WGC was recently discovered in the context of cosmic censorship \cite{Crisford:2017gsb}.  It was found that if you couple Einstein theory only to a $U(1)$ Maxwell theory, with sufficiently strong background electric field you can develop naked singularities, thus violating the cosmic censorship conjecture.  However, given conjecture 2 (that there are electrically charged states) and the WGC we deduce that there must also be light enough particles to be produced by such strong electric fields.  Taking this into account resolves the naked singularity and thus avoids the violation of cosmic censorship.

\subsection{Non-Supersymmetric AdS/CFT Holography belongs to the Swampland}

\emph{Non-supersymmetric AdS/CFT holography belongs to the swampland \cite{Ooguri:2016pdq}.}\\

Let us say immediately that this radical sounding claim is not saying that non-supersymmetric holography does not make sense in general. The claim is that non-supersymmetric AdS/CFT holography does not make sense, provided we have a finite number of particles. Indeed there can be versions of AdS/CFT, like in SYK \cite{Sachdev:1992fk,Kitaev} or higher spin Vasiliev theory \cite{Vasiliev:1995Dn}, with infinite towers of particles which do not lead to ordinary theories of gravity. We will not consider these cases. 

The motivation for this criterion is very simple. We typically get holography by putting branes next to each other in string theory and then by taking the near horizon limit. However, the problem is that if the branes are not supersymmetric, then the repulsion between branes wins over attraction due to WGC. In this case there is no way to keep the branes close to each other.  The refined WGC is simply saying that in the non-supersymmetric setup, those branes will repel and fly apart!

To illustrate this conjecture let us look at holography in the context of 2D CFT's. Consider a sigma model with target space given by symmetric products of $T^4$. The $AdS$ dual is known to be $AdS_3\times S^3\times T^4$.   More precisely in order to find a weak coupling holographic dual we need to blow up the singularity associated to coincident $T^4$'s.  This case is supersymmetric holography. In principle, one could think of doing the non-supersymmetric analog of exactly this construction by for example taking symmetric product of tori without fermions which does exist in the orbifold limit as a CFT. 
The problem arises when we try to find the $AdS$ dual, and for that we need to perform the blowup of the singularity.  But as can be readily checked the blow up modes are not marginal deformation of the CFT and so we cannot blow it up and we are stuck with the singularity and so we will not find a weak coupling non-supersymmetric $AdS$ dual\footnote{Another piece of evidence that supersymmetric holography with finite number of particles is in the swampland can be seen by examining the SYK model. In this case there is an infinite tower of massless states, so this conjecture does not apply. However,  
we can ask if there is some way to adjust the potential in the SYK model in order to truncate this infinite spectrum so that the model would violate the swampland conjectures. So far all attempts  have been unsuccessful.}.

The lack of stable non-supersymmetric $AdS$ geometries in a consistent theory of quantum gravity has interesting phenomenological implications related to neutrino physics \cite{Ooguri:2016pdq}.  This is related to the fact that depending on the neutrino mass types and ranges upon compactification to 3D one may obtain non-supersymmetric $AdS$ geometries in 3D \cite{ArkaniHamed:2007gg}. This places restrictions on neutrino mass types and ranges.  Interesting extensions of this have been recently considered in \cite{Ibanez:2017oqr}. The implications of these constraints on neutrino (and Higgs) physics upon compactification has also been studied recently in more detail in \cite{Hamada:2017yji}. 

\subsection{dS and the Swampland}
\label{sec:dS}
\emph{$dS$ space does not exist as a consistent quantum theory of gravity and it belongs to the swampland}.

We have seen that non-supersymmetric AdS/CFT holography lies in the swampland, while supersymmetric $AdS$ is of course possible. This can be summarized as:
\begin{equation}
\mbox{Non-SUSY is not allowed}\ \Longleftarrow\quad \Lambda<0\quad  \implies\ \mbox{SUSY is allowed}~,
\end{equation}
where $\Lambda$ is the cosmological constant.

One may wonder if the opposite situation happens for positive cosmological constant. In this case, we know that supersymmetric $dS$ does not exist, as it is impossible to define a supersymmetry algebra in a de Sitter spacetime \cite{Pilch:1984aw}. Could it then be possible to have non-supersymmetric $dS$ realized from string theory?
\begin{equation}
\mbox{SUSY is not allowed}\ \Longleftarrow\quad \Lambda>0\quad  \implies\ \mbox{Is Non-SUSY allowed?}
\end{equation}

Answering this question looks difficult for many reasons. There are typically two different ways in which we can get $dS$ in string theory:

\begin{enumerate}
\item Metastable $dS$, as in KKLT, LVS, etc. \cite{Kachru:2003aw,Cicoli:2008va}.
In this class of models the $dS$ vacuum is obtained from some uplift of a previously $AdS$ vacuum by introducing extra ingredients such as anti-D3-branes to make the cosmological constant positive. The scalar potential for the cosmological constant takes the form given in Figure \ref{KKLTfig}.

\begin{figure}[h!]
\centering
\includegraphics[scale=0.5]{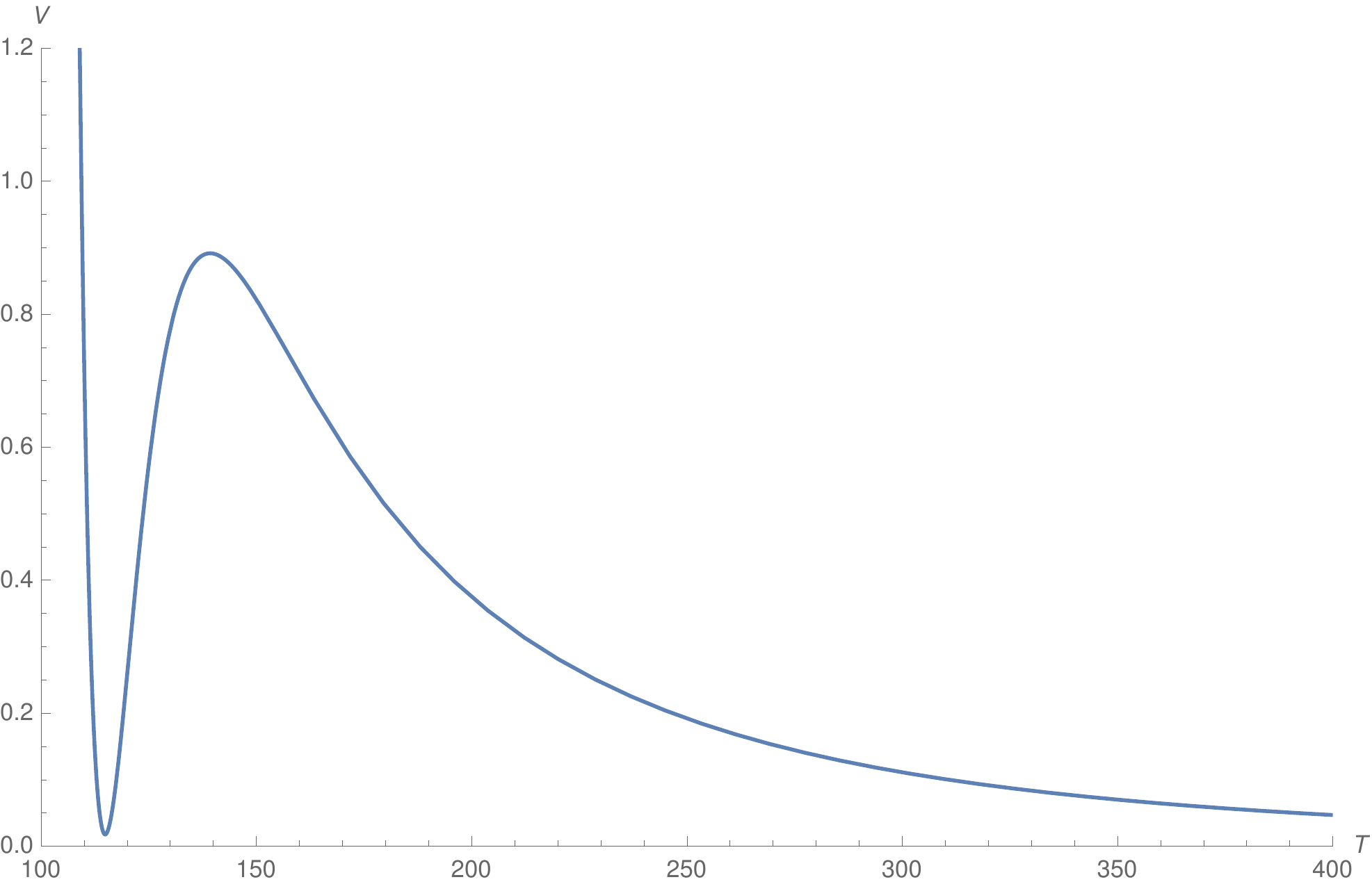}
\caption{The scalar potential for a metastable $dS$.}
\label{KKLTfig}
\end{figure}

\item Quitessence models. 
In this class of model the $dS$ vacuum is completely unstable, and slowly rolling down to the Minkowski case. The potential takes the form given in Figure \ref{Quintfig}.
\begin{figure}[h!]
\centering
\includegraphics[scale=0.5]{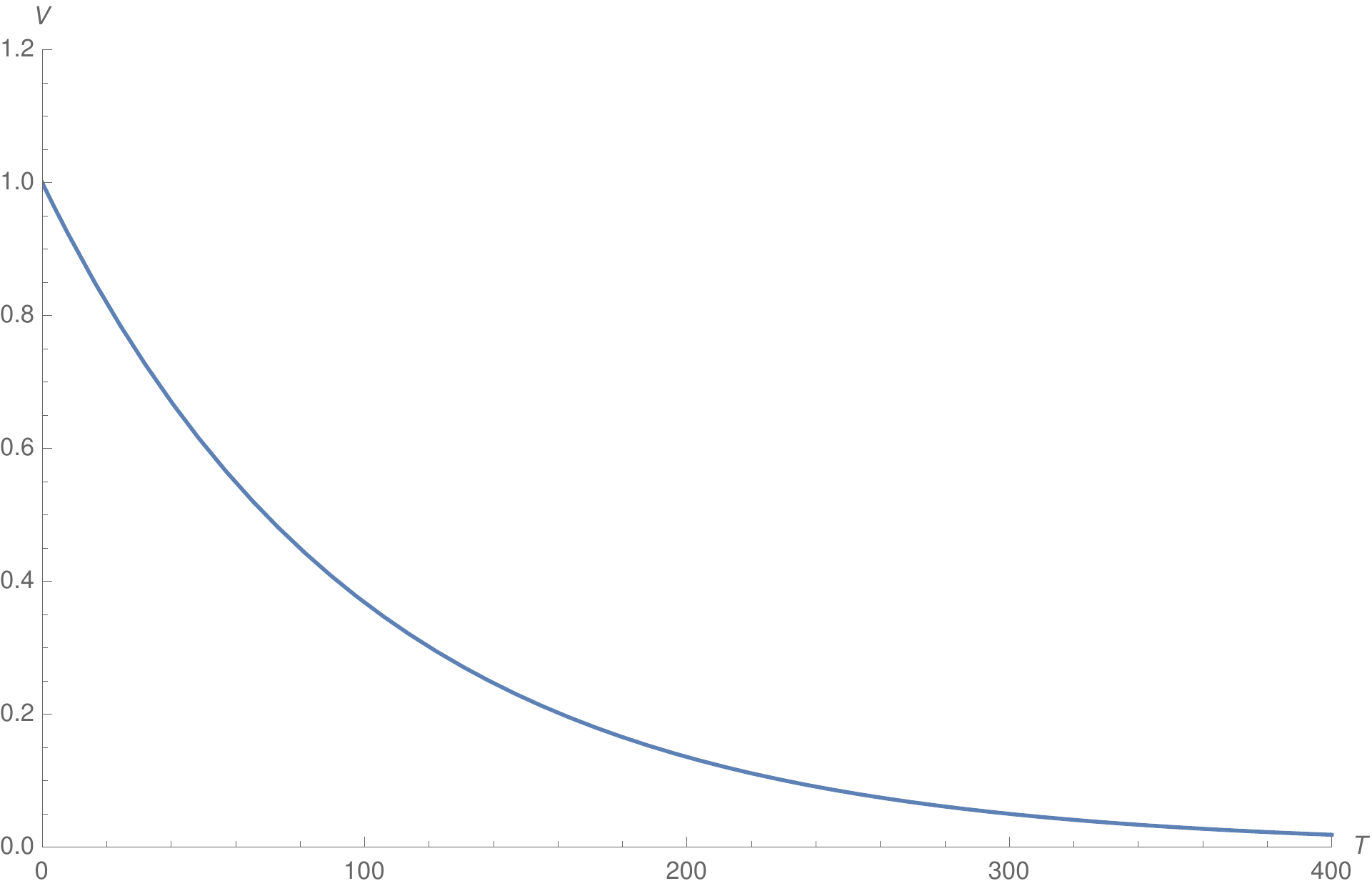}
\caption{The scalar potential for an unstable $dS$.}
\label{Quintfig}
\end{figure}

\end{enumerate}

We may conjecture that metastable $dS$ belongs to the swampland.  There are a number of no-go theorems for constructing $dS$ in string theory.  For example, an argument by Maldacena-Nunez \cite{Maldacena:2000mw} shows that in M-theory without strong curvature background $dS$ vacuum is not possible.   Of course this does not prove it for all backgrounds.  For exmaple,  it may possibly be avoided by considering orientifolds or higher stringy corrections.  Nevertheless, this and the many similar no-go theorems could as well be taken as mild evidence supporting this last Swampland Conjecture that $dS$ does not exist as part of any consistent quantum theory of gravity.
Even though we can write the EFT for $dS$ and quintessence, however it seems that all known examples from string theory which are computationally under control are of the quintessence type. 
For a recent discussion of this see  \cite{Sethi:2017phn}.
It is possible therefore that quintessence models are the only ones allowed in string theory.  This is the motivation for our last criteria for the swampland that 
$dS$ does not exist!  Arguments based on lack of holographic duals for $dS$ space have been advocated by L. Susskind\footnote{Private communication with C.V.} as another motivation for their lack of existence.

This is of course a conjecture, but it is physically well motivated. For example, we live in a universe right now which is about $14$ billion years old. The current value of the cosmological constant $\Lambda$ also defines a time scale, which is about $100$ billion years. Why the current age of the universe is so close to the Hubble scale? If we are in the metastable $dS$ case, this is quite hard to explain, since the metastable vacuum can be extremely long-lived whereas in the quintessence models it could more naturally be of the same scale. Maybe this points to the fact that we are always in a runaway situation, and there is no way to stabilize the cosmological constant to the present value.  If this is the case it should be observable in the near future by the measurement of $w\not=-1$ for the equation of state for the dark energy which we would conjecture should soon be found!\footnote{The current experimental bounds place $w=-1$ to within 5 percent \cite{Kumar:2012gr}.  So this is somewhat puzzling for the quintessence picture which suggests no particular reason for it to have this value.}

One can continue this logic by deducing that the corresponding scalar field responsible for quintessence should interact strongly with the dark sector.  This is because an extension of the WGC \cite{Palti:2017elp}\ would suggest that the scalar field has to couple stronger than gravity to some matter fields and we already know, by lack of violation of the equivalence principle in the visible matter sector, that this should be in the dark sector.  It would be interesting to find evidence for such a picture by finding apparent violation of equivalence principle in the dark sector due to the force generated by this scalar.



\section{Lecture 3: The Missing Corner}

One of the important promises of string theory is that it gives a UV complete description of quantum gravity including at the Planck scale. Despite being of primary interest, there is very little known about how to give a fundamental formulation of quantum gravity arising from string theory. One way we can try to study quantum gravity is by holographic duality.

Dualities are a crucial part of our understanding of string theory. In general, a duality relates two different descriptions of the same physical system -- each with different regimes of validity and utility. They relate a description with strong coupling (without a good perturbation series) to another with weak coupling (with a good perturbation series). This picture of dualities tells us that there is no physical system with two descriptions where both are weakly coupled. If there were, they would have to be exactly the same description as they would have to match every process order by order in perturbation. 

A good example of a duality in which we have full control over both sides of the theory is T-duality in string theory. This duality relates strings on $S^1_R$ to strings on $S^{1}_{R^\prime}$ where $R'=\ell_s^2/R$. In general, we only have a good description of the string states for $R>> \ell_s$. This is because we can only make sense of particle states on the circle if the radius is much larger than the Compton wavelength of the states. However, in the regime where $R<<\ell_s$, we can use the duality to map to the dual perturbative description where $R^\prime=\ell_s^2/R>>\ell_s$ and the perturbative modes are now the winding modes.   Just as position $x$ is related to Fourier transform of momentum states, we can define a new notion of position $x'$ suitable for winding states by: 
\be
|x\rangle=\sum_pe^{i px}|p\rangle\quad, \qquad  
|x'\rangle=\sum_we^{i wx'}|w\rangle~,
\ee
where $|p\rangle$ and $|w\rangle$ are the momentum and winding modes respectively, $x$ is periodic with period $2\pi R$, and $x'$ is periodic with period $2\pi R'$.

This mapping makes it clear that $x'$ is not a useful description of the theory for the case where $R>>\ell_s$ because a single wave packet is made up of a number of very massive winding modes.  On the other hand when $R<<\ell_s$ the momentum modes are very massive and $x$ becomes useless because we can hardly excite momentum modes. In this limit, the theory is more appropriately described by $x'$. The example of T-duality demonstrates the idea that in general, there is at most one useful description of a physical system for each point in parameter space and no description is singled out globally as the best description; sometimes there is no good description. We will call this philosophy the ``Democracy of Theories.''
Of course we can use either description in all ranges but the physics will look very complicated in terms of the wrong variables and we may have to define things which may look non-local with respect to that variable.

While dualities are an integral part of our understanding of string theory, the fact that there are many dual descriptions does not mean that string theory itself is an effective theory. String theory provides a good theory of quantum gravity that is perturbatively well defined to all orders and in fact is arguably the only complete quantum theory of gravity\footnote{One could make the argument that there are other theories of quantum gravity different from string theory, such as Vasiliev higher spin theories \cite{Vasiliev:1995Dn} or even loop quantum gravity \cite{Smolin:2004sx}. However, it is also possible that all these other theories could be attained as a special limit of string theory\cite{Dijkgraaf:2004te}.}. However, this restricts us to the regime of small $g_s$. 

In order to have a complete understanding of string theory, we need to have to go beyond perturbative description and find a full non-perturbative description. There have been many attempts to accomplish this. Some of these approaches include string field theory which has had varying degrees of success. The idea of string-field theory is to create a spacetime quantum field theory that would replicate all of the scattering amplitudes of full string theory. If such a theory were to exist we could hope to have multiple dual descriptions which would allow us to study the strong coupling limit of string theory and unify the different formulations. For a brief overview, see \cite{Siegel:1988yz,Rastelli:2005mz,deLacroix:2017lif}. While there has been tremendous successes along this direction, such as formulating open string theory as a Chern-Simons theory, there is still much that is unknown such as a good description of closed string field theory.  

\begin{figure}
\begin{center}
\includegraphics[scale=0.5]{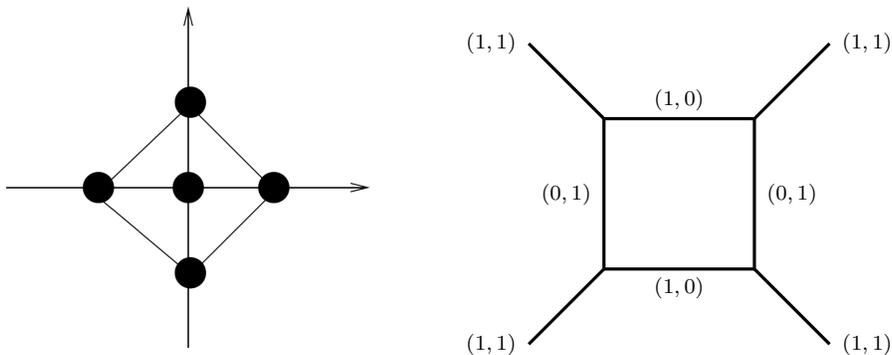}\qquad
\begin{tikzpicture}
\draw [-, very thick, black] (0,-1)--(0,1);
\draw [-, very thick, black] (2,-1)--(2,1);
\draw [-, very thick, black] (0,1)--(2,1);
\draw [-, very thick, black] (0,-1)--(2,-1);
\draw [-, very thick, black] (2,1)--(3,2);
\draw [-, very thick, black] (0,1)--(-1,2);
\draw [-, very thick, black] (0,-1)--(-1,-2);
\draw [-, very thick, black] (2,-1)--(3,-2);
\node (pqlab1) at (1,1.25) {$(1,0)$};
\node (pqlab1) at (1,-1.25) {$(1,0)$};
\node (pqlab1) at (2.5,0) {$(0,1)$};
\node (pqlab1) at (-0.5,0) {$(0,1)$};
\node (pqlab1) at (3.5,2) {$(1,1)$};
\node (pqlab1) at (-1.5,2) {$(1,1)$};
\node (pqlab1) at (3.5,-2) {$(1,1)$};
\node (pqlab1) at (-1.5,-2) {$(1,1)$};
\end{tikzpicture}

\caption{(Left) This is the toric diagram for the $CY^3$-fold given by the affine cone over $\IP^1\times \IP^1$. This is dual to a Type IIB brane web with given $(p,q)$ 5-branes (right)\cite{Leung:1997tw,Strominger:1995cz}. }
\label{fig:example}
\end{center}
\end{figure}

These facts seem to suggest that string theory is a complete theory even though we only have a perturbative understanding of it. To further illustrate the idea that string theory should be seen as a complete theory, let us consider Seiberg-Witten theory. In 1994, Seiberg and Witten solved for the low energy dynamics of  $\CN=2$ $SU(2)$ supersymmetric gauge theory   \cite{Seiberg:1994rs,Seiberg:1994aj}. They showed that the theory can be described by a $U(1)$ theory with coupling parameter $\tau(u)$ which is dependent on the vev of the vector multiplet scalar 
\be
u=\frac{1}{2}\left\langle\text{Tr }\phi^2\right\rangle~.
\ee
Since the $U(1)$ theory is not UV complete, just by field theory reasoning, it is clear that this must be a low energy effective theory for the $SU(2)$ SYM. However, we can UV complete both theories, the $\CN=2$ $SU(2)$ SYM theory and the $U(1)$ effective theory, by embedding them into string theory. This is evident from mirror symmetry where Type IIB on the $CY^3$-fold:
\be
\{v w = z+x^2-u+1/z\}\subset \IC^4~,
\ee
which describes a $U(1)$ theory is mirror dual to Type IIA with D-branes which describe an $SU(2)$ $\CN=2$ SYM theory \cite{Klemm:1996bj}.  In this duality, the worldsheet instantons of the Type IIA side are computed in the Type IIB mirror decription as period integrals of a holomorphic 3-form which reduces to the Seiberg-Witten solution \cite{Seiberg:1994rs,Seiberg:1994aj}. See Figure \ref{fig:example}. 
In this setup it is not correct to say that the Type IIB side is only an `effective' description of the physics and the Type IIA side is the `real' definition.  The Type IIA and Type IIB are both on the same footing in terms of defining a theory.  We thus see this as another example of `democracy of theories'.

Now we discuss a way to define quantum gravity which is called holography. But in order to do so, we will first need to take a brief historical detour. In the mid 1970s `t Hooft was studying the large $N$ limit of $SU(N)$ gauge theory \cite{tHooft:1973alw}. In taking the limit
\be
g_{YM}\to 0\quad,\qquad N\to \infty\quad,\qquad \lambda=g^2_{YM} N\text{  fixed ,}
\ee
the perturbation theory in $1/N$ becomes a sum over ribbon graphs which have the topology of Riemann surfaces. See \cite{Coleman,tHooft:2002ufq,Marino:2005sj}
 for a review. This is very reminiscent of the summation in string theory over worldsheet topologies. `t Hooft realized this and thought that at strong coupling the boundaries of the Riemann surfaces (or really the ribbon diagrams) could close up to form smooth closed surfaces without boundaries \cite{tHooft:1973alw}. Because of this, he suggested that perhaps {\it closed string theory would be a solution to strongly coupled Yang-Mills theory}. 

As it turns out, `t Hooft's intuition was correct. This can be exactly realized in string theory in the context of celebrated AdS/CFT correspondence \cite{Maldacena:1997re}. In this correspondence Type IIB string theory on $AdS^5\times S^5$ is dual to $\CN=4$ $SU(N)$ SYM theory on the boundary of $AdS_5$. Here the $N$ in the SYM theory relates to the size of the the $AdS_5$ space
\be
L_{AdS}^4=4\pi g_s N\alpha^{\prime 2}~.
\ee
This duality has been checked very rigorously in the large $N$ limit: string perturbation in $1/N$ matches to all orders in the expansion on the SYM theory. For more details see \cite{Aharony:1999ti}.

We can now ask the question if using this AdS/CFT correspondence gives a non-perturbative definition of string theory. The motivation for this is that we can give a non-perturbative definition of SYM theory, for example by lattice regularization, whereas the holographic quantum gravity dual theory in $AdS$ has no complete definition.  The fact that the CFT side, i.e. the non-perturbative definition of SYM, gives {\it in principle}, a non-perturbative definition of the $AdS$ side, is of course  true.  But this may be not very useful for deeper questions of quantum gravity.  In fact the regime that the gravity side is weakly coupled is big corresponds to when the SYM is strongly coupled.
In fact `t Hooft was trying to use string theory as a {\it solution} to the gauge theory question at strong coupling and not the other way around! 

This is analogous in the context of T-duality to defining the physics of a boson on $S^1$ using winding modes when the space is much larger than the string scale.
While the AdS/CFT duality can give us some very useful insights into the non-perturbative regime of string theory, it does not tell us directly how to describe it.   Some have argued that perhaps there is no direct definition of the $AdS$ side.  In a sense, gravity is always an `effective theory' rather than fundamental theory.  This is analogous to the example of the effective $U(1)$ theory in the SW example discussed above.  However, we saw in that case there is a complete string theory behind the would be effective $U(1)$ theory.  Moreover if there is no direct definition of $AdS$ side, the democracy of theories is violated: the  CFT side would be viewed as more fundamental than the $AdS$ side.  This is counter to the fundamental idea of a duality as well as to all the other known examples.

Despite this, the the AdS/CFT correspondence 
has given us a lot of insight into strongly coupled CFT by using the semi-classical gravitational picture. 
However, it is not a good tool for answering many questions we have about the bulk. This is because it is very hard to discuss bulk locality starting from the boundary theory, similar to how it is difficult to describe locality using winding modes in the T-duality example from before. This makes it difficult to answer some of the most interesting phenomena, such as what happens with black hole evaporation or with firewalls, and additionally suggests that the AdS/CFT correspondence should be used in order to try to understand the CFT side, rather then attempting to use the CFT side in order to define and study quantum gravity in $AdS$.   

After this very long introduction to the problem, 
we find ourself back at the beginning: 
we want to know fundamentally, what is quantum gravity? It should describe the quantum fluctuations of the metric. From a brief analysis of the standard Einstein-Hilbert action, we see that fluctuations of the metric at the Planck scale should become very violent, leading to potential changes in the topology of the spacetime \cite{Wheeler,Hawking:1979zw}. This leads naturally to the idea that quantum gravity should be equivalent to summing over all spacetime topologies and geometries:
\be
Z_{QG}\sim  \text{ }\sum_{\text{top. and geo.}} e^{-S}\text{ }~.
\ee

In general we have no idea about what description will lead to the correct sum over geometries and topologies. We only do know that there should be some mechanism that washes out the Planck scale fluctuations to produce a smooth space at lower energies. It seems that this description must come from some new fundamental 
principle, rather than from some duality such as mirror symmetry or AdS/CFT. This lack of knowledge of describing the gravity side quantum mechanically is ``the missing corner" in our understanding of string theory. 

\subsection{Introduction to Topological String Theory}

One case where we have special insight on how to give a non-perturbative description of quantum gravity is in topological string theory. This can be thought of as a sort of toy model of string theory that was introduced by Witten \cite{Witten:1988ze,Witten:1988xj,Witten:1988xi}. Topological string theory can be seen as a restriction to a special, supersymmetric subspace of the Hilbert space of full string theory \cite{Bershadsky:1993ta,Bershadsky:1993cx}.  A review of basic aspects of topological strings is given in Appendix \ref{app:B}.

As discussed there, we have two types of topological strings A-model or B-model, which we will take to be on a Calabi-Yau 3-fold $X$. The  A-model only depends on Kahler structure of $X$ and B-model
depends only on the complex structure.  Moreover they are related to one another by mirror symmetry.  The path-integral for the A-model is restricted to be on holomorphic maps to CY 3-folds.  The partition function is given as
\be Z={\rm exp}\left[ \sum_g F_g(t) g_s^{2g-2}\right]~,\ee
where $F_g(t)=\sum_d n_g^d \ {\rm exp}(-d\cdot t)$ is the contribution of the genus $g$ worldsheet to the free energy, $t$ denotes the Kahler parameter, $n_g^d$ is the `number' of curves of genus $g$ with degree $d$ in $X$, and $g_s$ is the string coupling constant.  The B-model is the mirror of this computation and $t$ gets mapped to complex deformation parameters.  Moreover $F_0$ on the B-model side is captured by the period integrals of 3-forms on $X$.

We can also make sense of open topological string theory by considering worldsheets that have boundary components. In this case the boundary components map to branes in the target space. \\

\textbf{\underline{A-model}: }
In the topological A-model, these boundary components will map to Lagrangian submanifolds of the target space. This is because the map $\phi^i:\Sigma\to X$ is independent of the complex structure and hence the boundary component must also be invariant. 

Consider a stack of $N$ D-branes wrapped on $M\subset X$ a Lagrangian submanifold. The local structure near $M$ is of the form $T^\ast M$ so that in a local patch $\CU\subset M$ with local coordinates $q_a$, then the K\"ahler form is of the form
\be
J=\sum_a dp_a\wedge dq_a~,
\ee
where $p_a$ are the fiber coordinates on the trivialized $T^\ast \CU\cong \IR^3\times \CU\subset T^\ast M$. A string field theory computation following \cite{Witten:1992fb} shows that the branes will induce an analytically continued Chern-Simons theory on the worldvolume of the D-branes given by 
\be
S_{brane}=\left(\frac{1}{2g_s}\right)\int_M\text{ Tr}\left(A\wedge F+\frac{2}{3}A\wedge A\wedge A\right)~,
\ee
with gauge group $G=U(N)$. Here we say analytically continued Chern-Simons theory because we have that the level is generically non-integer. \\

\textbf{\underline{B-model}: }
In the topological B-model, these boundary components will map to holomorphic submanifolds of the target space. Consequently this will induce holomorphic Chern-Simons theory on the worldvolume of the wrapped D-branes
\be
S=\left(\frac{1}{2g_s}\right)\int_X \Omega\wedge\text{ Tr}\left(A\wedge \bar\partial A+\frac{2}{3}A\wedge A\wedge A\right)~.
\ee
Consider compactifying Type IIB string theory on the $CY^3$-fold $X$
\be
X\equiv\{uv+y^2+W'(x)^2=0\}\subset \IC^4~,
\ee
For each critical point of $W(x)$: $W'(x)=0$ we get locally a conifold geometry which can be reolved.  We can wrap branes around it.  This 
leads to a 4D $\CN=2$ theory which is broken to an $\CN=1$ by giving a superpotential to the adjoint field Tr $W(\Phi)$  \cite{Cachazo:2001jy,Klemm:1996bj,Dijkgraaf:2002fc,Dijkgraaf:2009pc,Cheng:2010yw}. 

If we consider the topological B-model on $X$ we again see that there will be branes wrapping the holomorphic 2-cycles given by the degenerate 2-spheres given by blowing up
\be
uv+y^2=0~.
\ee
The theory in this case is described by a matrix model with action given by \footnote{It is interesting to note that these matrix models (with suitable choices of $W$) are dual to Liouville theory on the Riemann surface given by $\Sigma_{SW}$ above. As it turns out, this theory is exactly the 2D CFT that describes the vertex operators corresponding to brane insertions and further is 2D CFT associated to the AGT correspondence which describes the physics of the corresponding four-dimensional theory of class $\CS$ \cite{Alday:2009aq,Dijkgraaf:2009pc,Cheng:2010yw}.}
\be
S=\frac{1}{g_s}\text{Tr }W(\Phi)~.
\ee

%

\subsection{Large $N$ Holography in Topological String Theory}

Topological string theory is a very powerful tool and a good first step towards understanding string theory in its full generality. In addition to giving tools for studying exact quantities in 4D theories, it also gives a clear manifestation of large $N$ duality and holography.

One of the key features of D-branes in string theory is that they source $p$-form flux. In addition, they can often be exchanged for a different background geometry supported by their sourced flux as in AdS/CFT \cite{Maldacena:1997re}. In topological string theory we would expect a similar behavior. In both the topological A- and the B-model, there is only a single $p$-form: the K\"ahler 2-form and holomorphic 3-form respectively. This means that the branes in each theory must support the respective $p$-form field of the theory. In the A-model, D-branes wrap a Lagrangian 3-cycle $L$ such that we can link it with a homologically trivial 2-cycle $C$ (since together they have co-dimension 1 in $X$) such that $C$ is non-trivial in $X\backslash L$. Since $C$ is a trivial in the absence of branes
\be
\int_C k=0~,
\ee
since $dk=0$ by nature of being a K\"ahler form. However, once we wrap branes on $L$, this result changes to count the flux of the D-branes
\be
\int_C k= Ng_s~,
\ee
where $N$ is the number of D-branes wrapped on $L$. 

Similarly in the B-model, the D-branes are wrapped on holomorphic 2-cycles. Following the same argument, we find that there is a homologically trivial 3-cycle $Y$ linking any holomorphic 2-cycle $M$ such that if we wrap $N$ D-branes on $M$, then the integral
\be
\int_Y \Omega=N g_s~. 
\ee

Now we discuss how the large $N$ duality works in this context and how it relates to geometric transitions \cite{Gopakumar:1998ki}. Consider the topological A-model with a real codimension 3 Lagrangian submanifold $M^3\subset CY^3$. Locally $CY^3$ looks like $M^3$ times the normal direction in $CY^3$ which is a cotangent space and can be written as $CY^3\sim T^\ast M^3$. Now wrap $N$ D-branes on $M^3$. 
 Now if we wrap $N$ D-branes on $M^3$, then we have open topological strings ending on the D-branes.  As already discussed, the effective theory on these branes is given by complex Chern-Simons theory $CS_k(U(N);M^3)$.

However, these branes back-react on the geometry. By integrating the K\"ahler class over an $S^2$ surrounding $M^3$ in the fiber of the cotangent bundle we find
\be
\int_{S^2} k=N g_s~.
\ee
This means that we can interpret D-branes as sourcing the volume for the $S^2$. In other words the D-branes can be replaced by giving finite size to this $S^2$.  The bigger the $N$ is the bigger this $S^2$ becomes.  This is the content of large $N$ duality in A-model topological string \cite{Gopakumar:1998ki}.
As an example consider the topological A-model on the conifold $CY^3=T^\ast S^3$ with $N$ D-branes wrapped on the base $S^3$. 

\begin{figure}
\begin{center}
\includegraphics[trim=1cm 23cm 4cm 1cm,clip]{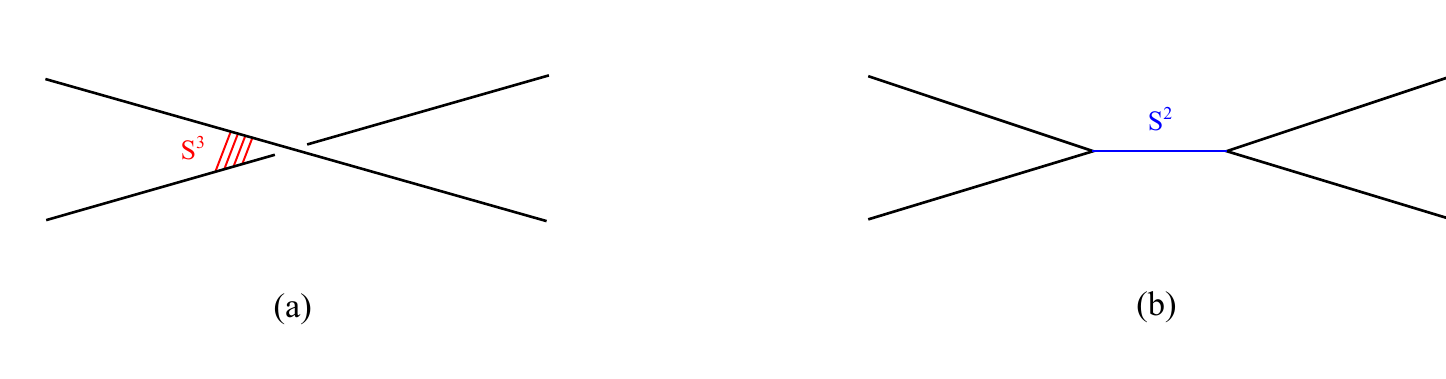}
\caption{Here the holographic duality replaces an $S^3$ resolution (a) with an $\IP^1$ resolution (b) of the singular conifold.}
\label{fig:conifold}
\end{center}
\end{figure}

The normal direction to the base $S^3$ in $T^\ast S^3$ is given by $\IR^3$ so that the boundary $\partial(T^\ast S^3)$ is given by $S^2\times S^3$ at infinity. The $S^2$ links with the $S^3$ and so the D-branes on it give is a finite size:  $\int_{S^2}k=N g_s$.  This means we will have a geometric transition where $S^3$ shrinks and $S^2$ has now a finite size leading to a geometry $\CO(-1)\oplus \CO(-1)\to \IP^1$ where $S^2=\IP^1$ at the zero section of this bundle.   So we end up with topological A-model on the resolved conifold without any branes, but with finite size $\IP^1$.   The geometric transition underlying this holographic duality is exactly the physical manifestation of the conifold transition \cite{Candelas:1987kf} as in Figure \ref{fig:conifold} and Figure \ref{fig:conRes}.  This large $N$ duality can be checked by computing both sides independently.  The partition function of closed string side which involves considering holomoprhic maps to the resolved conifold agrees to all orders in the perturbative expansion with the Chern-Simons perturbative expansion for $U(N)$ on $S^3$ \cite{Gopakumar:1998ki}. Therefore, this holographic duality is indeed true. 
And as in the conifold transition, this holographic duality can be generalized to all toric geometries sitting inside $CY^3$-folds by gluing together building blocks by using the technology of the topological vertex  \cite{Aganagic:2003Db}.  

\subsection{Missing Corner for $\mathbb{C}^3$}

Now we can ask if there is any definition of the theory on the closed string side which is, from the target space point of view, a non-perturbative theory of gravity? 
 We will show that this is indeed the case and show how to recover the full partition function in topological string theory in yet another way.
In other words, we fill the missing corner of what the quantum gravity means in this topological setup. We will proceed by computing the example of $\IC^3$ since the techniques used generalize to other $CY^3$-folds \cite{Iqbal:2003Ds}.

Consider the topological A-model on $\IC^3$. Since $\IC^3$ is non-compact, all maps $X:\Sigma\to\IC^3$ are constant maps -- i.e. they map to a point in $\IC^3$ which we can without loss of generality take to be the origin. 
Using various arguments using topological string dualities and Chern-Simons theory \cite{Bershadsky:1993cx,Gopakumar:1998ii,Gopakumar:1998jq,Faber:2000ma,Okounkov:2003sp}, the partition function of this theory has been computed to be
\be
Z(g_s)=exp\left\{\sum_g g_s^{2g-2} \int_{\CM(\Sigma_g)} c_{g-1}^3\right\}=\frac{1}{\prod_{n=1}^\infty (1-q^n)^n}\quad,\qquad q=e^{-g_s}~,\label{eq:rawint}
\ee
where $\CM_g$ is the moduli space of Riemann surfaces with genus $g$, and $c_{g-1}$ is the $(g-1)^{th}$ Chern class of the Hodge bundle $\CH\to \CM_g$ over the moduli space\footnote{The Hodge bundle is the line bundle (equivalently a $U(1)$ gauge bundle) associated to the top holomorphic form $\Omega$ (which has a phase redundancy). This bundle has a metric whose associated Kahler function is
\be
h=i \int \Omega\wedge \bar\Omega~,
\ee
which has a compatible connection with a generically non-trivial curvature $F$. Chern classes are differential forms given by wedge products of the curvature of a bundle that encode topological data of the bundle.}.

Now we can ask if there is a target space or quantum gravitational formulation of this result where we sum over all possible geometries and topologies as we would expect from a theory of quantum gravity? As it turns out there is. By using a string field theory computation using the Batalin-Vilkovisky formalism \cite{Batalin:1981jr,Batalin:1984jr}, the classical action for the A-model is given by \cite{Bershadsky:1994sr}
\be
S=\frac{1}{g_s^2}\int_{CY^3}k\wedge k \wedge k~.
\ee
This can be viewed as
\be
S_{cl}=\frac{1}{g_s^2}Vol(CY^3)=\frac{1}{g_s^2}\int_{CY^3}k\wedge k\wedge k~
\ee
which we can think about as coming from a cosmological constant term. 
Now we want to try to reinterpret the result of \ref{eq:rawint} as a sum over changing spacetime topologies. Note that we are summing over the moduli space of K\"ahler classes of the manifold $X$. A key feature of K\"ahler forms is that they are closed forms
\be
dk=0~.
\ee
In a sense, we can then interpret them as the curvature of a line bundle -- the field strength of a $U(1)$ gauge bundle which is classified by its first Chern class $c_1=k$. Now in the sum over these line bundles we have to integrate over the non-trivial classes such that 
\be
\int_{M_2}k \neq0~.
\ee
However, as it turns out, in order to reproduce the results of \ref{eq:rawint}, we must implement a quantization condition 
\be
\int_{M_2} k=g_s N\quad,\quad N\in \IZ^+~.\label{eq:kahlerint}
\ee
This quantization of the  K\"ahler form implies that the $k$ form should be in the class
\be
[k]=g_s\delta^{(3)}(L)~,
\ee
where the D-branes are wrapping a Lagrangian 3-cycle $L$ and that spacetime geometry fluctuations should be sourced purely by D-branes. 
 This suggests that we should rather take $k=g_s F$ for $F\in H^{(1,1)}(CY^3;\IZ)$.   Here we interpret $F$ as curvature of a $U(1)$ gauge field.
 In this case we now have that 
\be
S=g_s\int_{CY^3} F\wedge F\wedge F~.
\ee

Now we are summing over K\"ahler classes with singularities. Without loss of generality, we can take these singularities to be at the origin -- arising from D-branes wrapping a collapsed 3-cycle -- giving rise to the non-trivial integral \ref{eq:kahlerint}. In the line bundle interpretation, this corresponds to summing over singular line bundles localized over the origin since the curvature is only non-trivial there.

By performing blow ups of this geometric singularity at origin, the singular line bundles are replaced with smooth line bundles that have non-trivial curvature on the blown up geometry. By blowing up a sufficient number of times, we can in fact make any line bundle smooth so that the curvature has a single unit of charge for each blown up $\IC\IP^1$. In this way we can translate the sum over singular K\"ahler classes to actual changes in spacetime topology \cite{Iqbal:2003Ds}. 

Now we can rewrite  the action as
\be
S=g_s\int_{CY^3} ch_3~,
\ee
where $ch_3$ is the third Chern character of a line bundle over the different components of the blown up geometry. 
Summing over the line bundles (that is $U(1)$ gauge fluxes) or equivalently the blown up geometry amounts to counting the number of sections of these line bundles which can be realized as the number of terms in the polynomial 
\begin{figure}
\begin{center}
\includegraphics[scale=0.8,trim=1cm 23.5cm 2cm 3.1cm,clip]{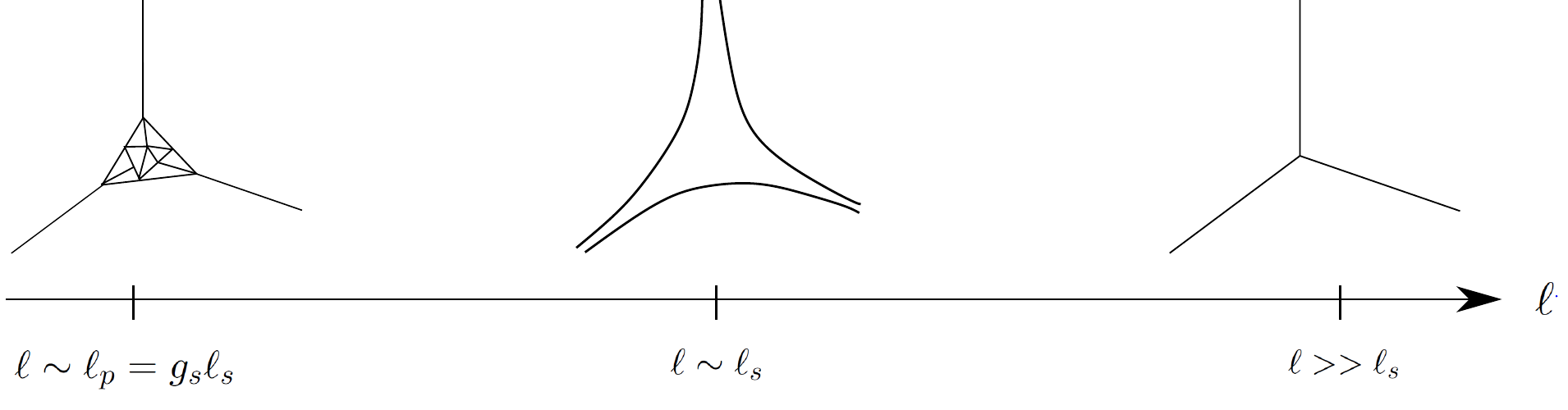}
\caption{Here we demonstrate the changing spacetime geometry depending on energy scale $\ell$. When $\ell>>\ell_s$, spacetime is that of classical geometry, $\ell\sim \ell_s$ is a smooth quantum geometry, and $\ell\sim \ell_p=g_s\ell_s$ is the ``quantum foam" with violently changing topology given by fluctuating blow ups.}
\label{fig:SG}
\end{center}
\end{figure}
\be
\sum_{n_i} a_{n_1,n_2,n_3}z_1^{n_1}z_2^{n_2}z_3^{n_3} \quad,
\ee
where the non-vanishing coefficients $a_{n_1,n_2,n_3}$ in the sum are constrained based on the blown up geometry. 

Specifically viewing the ${n_1,n_2,n_3}$ as giving an octant of a lattice $\IZ_+^3$, then we can blow up by taking 3D Young diagrams and removing points starting from the corner near $(0,0,0)$.  Then we would have the restriction that the non-vanishing coefficients of $a$ correspond to $(n_1,n_2,n_3)\in \IZ_+^3$ in the complement of the deleted set of points \cite{Iqbal:2003Ds}. This physically corresponds to a sum over all of the ways in which the flux can be ``distributed'' among different blown up geometries. 
So deleting no points gives a contribution $1$.  Deleting the origin give a contribution of $q=e^{-g_s}$.  Deleting the origin and one of the three points next to it, gives the contribution of $3q^2$, etc.  When we take all of these contributions and sum them we end up with
\be
Z(g_s)=\frac{1}{\prod_{n=1}^\infty (1-q^n)^n} =1+q+3q^2+...~,
\ee
thus reproducing the perturbative closed string answer in a rather novel way.  This is the quantum gravitational foam realization of the same partition function filling the missing corner in the description of the quantum gravity side.  The quantum foam gives a different description of the geometry depending on which scale we consider.  This scale dependent view of spacetime is shown in Figure \ref{fig:SG}.  This gives a satisfactory realization of how smooth geometry emerges in the limit of $g_s<<1$, when we look at scales much bigger than the Planck scale of $g_s l_s$.


We have now seen that using topological string theory as a toy model of full string theory provides many promising results. Besides giving tools to study four dimensional quantum field theories, it also has many other properties we know to hold in full string theory such as AdS/CFT type holography. We have seen that topological string theory suggests that in the full string theory there may be an independent complete definition of the gravity side, which will in particular include a sum over spacetime topologies but still give rise to a smooth spacetime geometry at large scales.   Topological string theory thus strongly suggests that indeed there is a missing corner in our understanding of quantum gravity. 

\section*{Acknowledgements} 

The authors would like to thank the organizers of the Theoretical Advanced Study Institute summer program and the University of Coloardo Boulder. 
TDB and CV would additionally like to thank the Simons Foundation and the organizers of the 2017 Simons Summer Workshop for hospitality while writing this paper. TDB is supported by the U.S. Department of Energy under grant DOE-SC0010008 to Rutgers University. The work of F.C. is supported through a fellowship of the international programme ``La Caixa-Severo Ochoa", and the grants FPA2015-65480-P (MINECO/FEDER EU) of the “Centro de Excelencia Severo Ochoa” Programme, and the ERC Advanced Grant SPLE under contract ERC-2012-ADG-20120216-320421.   The research of CV is supported in
part by NSF grant PHY-1067976.

\appendix

\section{Branes Probing Singularities}
\label{app:A}

Another important facet of understanding the role of branes and singularities in string theory is how they interact when a brane probes a singularity (when a brane transversely intersects a singularity). There has been a large amount of research on this in the past twenty years \cite{Douglas:1996sw,Feng:2000mi,Feng:2005gw,Franco:2005sm,Yamazaki:2008bt} and has recently been revived as a possible extension of the story of theories of class $\CS$ 
\cite{Gaiotto:2015usa,Feng:2005gw,Franco:2005sm,Yamazaki:2008bt}. 

We will first consider the work of \cite{Douglas:1996sw} on branes probing orbifold singularities where a stack of $N+1$ D3-branes probes an orbifold singularity $\IC^2/\IZ_n$. Consider a spacetime $M_6\times \IC^2/\IZ_n$ where the D3-branes are transverse to and localized at the origin of $\IC^2/\IZ_n$. The worldvolume theory of a free stack of D3-branes is given by four dimensional $U(N)$ $\CN=4$ SYM theory to leading order. The presence of the $\IC^2/\IZ_n$ singularity breaks SUSY by half since rotational symmetry is no longer a symmetry along this direction, thus reducing the $R$-symmetry group, resulting in a four-dimensional $\CN=2$ theory.

Intuitively, since the fields living on the D3-branes come from oscillations of strings stretching in the perpendicular directions, they are acted on by the $\IZ_n$ action. This means that they generically decompose as irreducible representations of the $\IZ_n$ action so that a generic field $\Phi$ decomposes as
\be
\Phi(x)=\bigoplus_{i=1}^{n-1}\Phi^{(i)}(x)\rho_i~,
\ee
where the $\rho_i$ are irreducible representations. This means that we now have a collection of fields $\Phi^{(i)}(x)$ by projecting onto the different irreducible representations. 

More precisely, the $\IZ_n$ action acts on the fields by global gauge transformations coupled with an R-symmetry transformation. This means that the gauge group is broken down to the commutant of $\IZ_n$. Generically this means that the gauge symmetry is broken
\be
U(N)\to \prod_i U(N_i)\quad,\qquad \sum_i N_i=N~,
\ee
where the $\{N_i\}$ are determined by the embedding of $\iota:\IZ_n\hookrightarrow U(N)$. Under this embedding, the $\CN=4$ vector multiplet is broken into a sum over vector multiplets with gauge group $U(N_i)$ and bifundamental hypermultiplets of the $U(N_i)\times U(N_{i+1})$ representation. 

We can additionally couple the D3-branes to D7-branes so that the D3-D7 strings introduce fundamental hypermultiplets in the worldvolume theory of the D3-branes \cite{Douglas:1996sw}. These fields will similarly decompose as a sum over $\IZ_n$ representations depending on the embedding of the $\iota_f:\IZ_n\hookrightarrow G_F$ where $G_F$ is the flavor symmetry. This will lead to a collection $F_i$ of fundamental hypermultiplets coupled to each $U(N_i)$ vector multiplet. This leads to a cyclic quiver with fundamental hypermultiplets as displayed in the quiver below:

\begin{center}
\begin{tikzpicture}[node distance=1.3cm,
cnode/.style={circle,draw,thick,text centered},snode/.style={rectangle,draw,thick,text centered,minimum size=8mm}]
\node [cnode] (1) {$N_1$};
\node (2) [cnode,below right of=1,xshift=0.5cm] {$N_2$};
\node (3) [cnode,below of=2,yshift=-0.3cm] {$N_3$};
\node (4) [cnode,below left of=3,xshift=-0.5cm] {$N_4$};
\node (5) [cnode,above left of=4,xshift=-0.5cm] {$N_5$};
\node (6) [cnode, above of=5,yshift=0.3cm] {$N_N$};
\node (7) [snode, above of=1] {$F_1$};
\node (8) [snode, above right of=2] {$F_2$};
\node (9) [snode, below right of=3] {$F_3$};
\node (10) [snode, below of=4] {$F_4$};
\node (11) [snode,below left of=5] {$F_5$};
\node (12) [snode,above left of=6] {$F_N$};
\draw[-] (1) -- (2);
\draw[-] (3) -- (2);
\draw[-] (3) -- (4);
\draw[-] (4) -- (5);
\draw[dashed] (5) -- (6);
\draw[-] (6) -- (1);
\draw[-] (1) -- (7);
\draw[-] (2) -- (8);
\draw[-] (3) -- (9);
\draw[-] (4) -- (10);
\draw[-] (5) -- (11);
\draw[-] (6) -- (12);
\end{tikzpicture}
\end{center}
In general, there are other types of orbifold singularities which are of the form $\IC^2/\Gamma_\fg$. These theories will generally lead to quivers which are are of the form of the Dynkin diagram of the associated affine Lie algebra $\hat\fg$. See \cite{Douglas:1996sw} for more details.

We can additionally consider branes probing more complicated singularities. We will consider a wide class which will illustrate the general method for analyzing generic singularities of a Calabi-Yau manifold. Consider Type IIB on a generic $CY^3$-fold $X$ with singularities. The allowed singularities for these manifolds are toric -- that is locally they locally look like a toric singularity $Y\times \IC$ where $Y$ is a toric variety with degenerating $T^2$ fibers \cite{Feng:2000mi}. Now consider a stack of D3-branes probing this singularity. As in the case of orbifold singularities above, the structure of the singularity restricts the fluctuations of the D3-branes and hence modify their worldvolume theory. Note that since we are considering Type IIB on a $CY^3$-fold with transverse D3-branes, we are generically studying a four-dimensional $\CN=1$ theory.

\begin{figure}[t]
\centering
\includegraphics[scale=0.9,clip,trim=1cm 23.5cm 4.5cm 1cm]{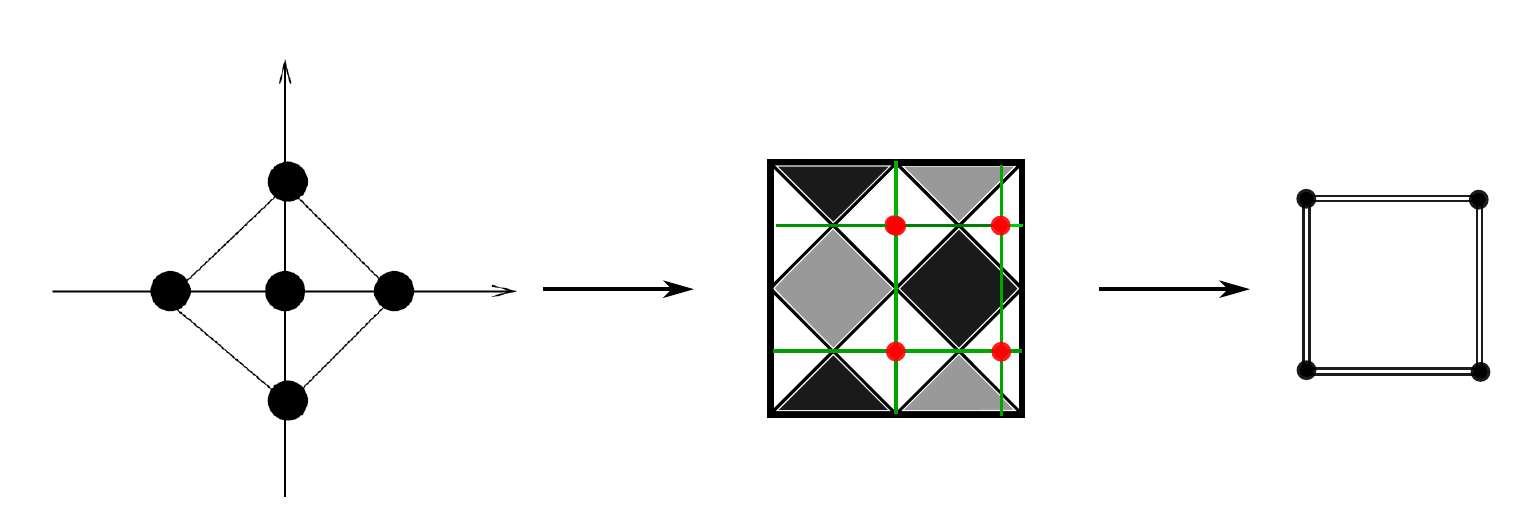}
\caption{This figure shows how to pass from toric diagram (left) to brane tiling (center) to quiver gauge theory (right).  In going from the toric diagram to the brane tiling, the normal of the external legs of the toric diagram give the angles of the cycles on the torus where the NS5-branes intersect the stack of D5-branes. This leads to the tiling in the center where the white are stacks of branes with charge $(N,0)$, black are stacks with $(N,+1)$ charge and grey are stacks with $(N,-1)$ charge. In going from the brane tiling to the quiver, we associate a quiver node with each white tile and then attach bifundamental hypermultiplets between nodes corresponding to adjacent faces.}
\label{fig:BraneTilingEx}
\end{figure}

Consider a stack of $N$ D3-branes probing a toric singularity. The worldvolume theory is given by a quiver gauge theory that arises from what is called a brane tiling. Recall that the toric singularity can be described by a toric diagram as in Figure \ref{fig:toric} where the dual graph describes where the different cycles of the $T^2$ fiber degenerate.  The idea of a brane tiling is that we have $N$ D5-branes wrapping a $T^2$ which is partitioned into ``tiles'' by transversely intersecting NS5-branes whose angle on the torus is determined by the degenerate cycle of the external components of the toric diagram. These pieces together form a tiling of the torus where each tile is a stack of $(N,\pm1)$ or $(N,0)$ 5-branes. This tiling creates a 3-colored torus where no two adjacent tiles are of the same color. Then the stacks of $(N,0)$-branes give rise to a four-dimensional $\CN=1$ gauge theory since the combination of NS- and D-type boundary conditions on the $(N,\pm1)$ 5-brane stacks freeze out massless degrees of freedom. Then we get bifundamental chiral multiplets by diagonally touching stacks of $(N,0)$ branes which results in a four-dimensional $\CN=1 $ quiver gauge theory. As an example, see Figure \ref{fig:BraneTilingEx}

\begin{figure}[t]
\centering
\includegraphics[trim=1cm 24cm 4cm 1cm,clip]{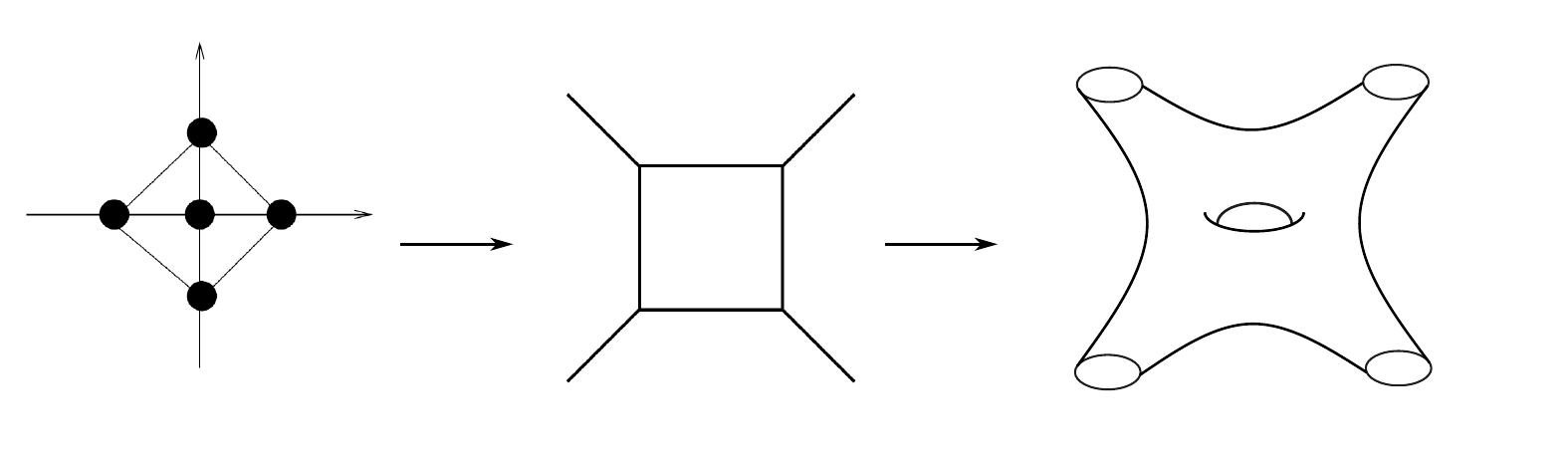}
\caption{This figure shows how the Riemann surface $\Sigma_0$ (right) is related to the data of the toric diagram (left). The Riemann surface associated to the brane tiling is the thickened dual graph to the toric diagram.}
\label{fig:TilingFlow}
\end{figure}

This brane tiling arises physically as follows. Consider the mirror dual of the stack of $N$ D3-branes probing the toric singularity. This is given by a stack of $N$ D6-branes wrapping non-trivial 3-cycles in the dual $CY^3$-fold given by 
\be
uv +P(x,y)=0\quad,\qquad P(x)=\sum_{(p,q)\in Q}c_{(p,q)} x^p y^q~,
\ee
where $Q$ are the coordinates of the toric diagram and $c_{(p,q)}$ are unfixed coefficients describing the moduli of the $CY^3$-fold. We can more conveniently write this as a double fibration over $\IC_W$, which we will call the $W$-plane, 
\be
\pi:\{P(x,y)=w,~uv=w\}\to w\in \IC_W~.
\ee
This is a toric $CY^3$-fold where the three $S^1$ fibers are given by the action
\be
x\to \alpha x\quad,\qquad y\to \beta y\quad,\qquad u/|u|\to \gamma u/|u|~.
\ee
These fibers degenerate at $w=0\in \IC_W$ where the $\gamma$-fiber degenerates. Here the $(x,y)$-fiber of is given by a Riemann surface $\Sigma_0$ which is defined by 
\be
\Sigma_0=\{P(x,y)=0\}~.
\ee
See Figure \ref{fig:TilingFlow} for more details on how this is related to the toric diagram

Therefore, we can view the $CY^3$-fold as a $\Sigma_w$-fibration. This Riemann surface has an embedded $S^1_s\hookrightarrow \Sigma_{w_s}$ that degenerates at points $w_s\in \IC_W$ where
\begin{figure}[b]
\centering
\includegraphics[trim=3.5cm 18cm 7cm 5.5cm,clip]{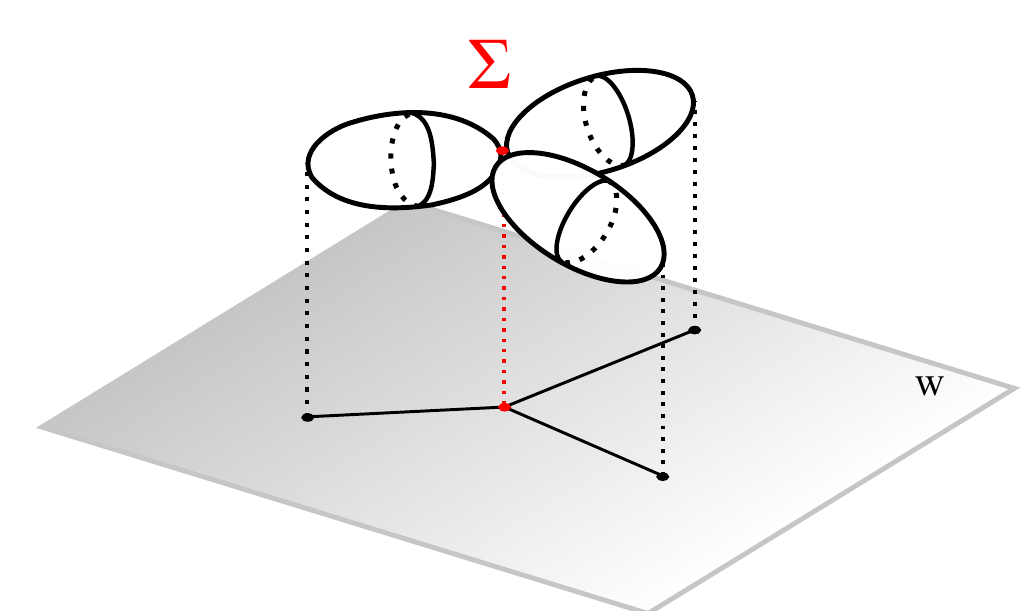}
\caption{This figure illustrates the mirror dual of D3-branes probing a toric singularity of a $CY^3$-fold. Under mirror duality, the D3-branes become D6-branes wrapping 3-cycles which are shown here as fibered over the $W$-plane intersecting the $\Sigma$ fibered over the origin.}
\label{fig:WPlane}
\end{figure}
\be
w_s=P(x_s,y_s)\quad,\qquad \partial_x P\Big{|}_{(x_s,y_s)}=\partial_yP\Big{|}_{(x_s,y_s)}=0~.
\ee
The D6-branes, which are the mirror image of the D3-branes, wrap the closed 3-cycles that are formed by the pre-image of a $T^2\subset T^3$ fiber over the straight line in the $W$-plane from the $w_{s}$ to the origin. Here the $T^2$ fibers are locally given by $S^1_\gamma S^1_{s}$'s where $S^1_{s_i}$ degenerates at $w_{s}\in \IC_W$. See Figure \ref{fig:WPlane}.

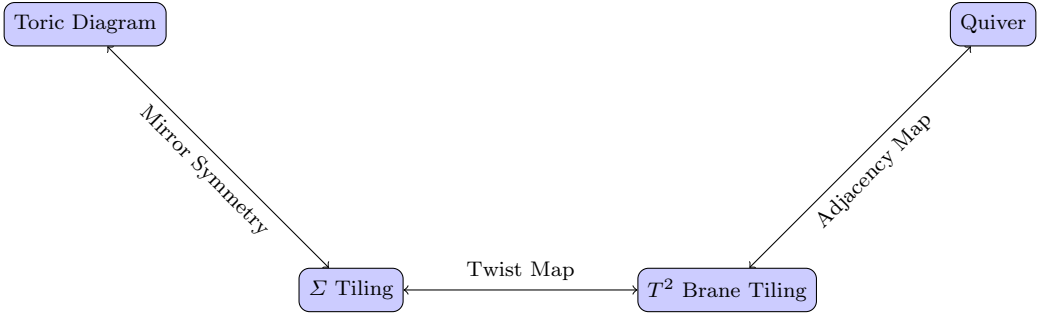
\begin{figure}[t]
\centering
\begin{tikzpicture}[node distance = 5cm, auto,<->]
\node [block] (toric) {Toric Diagram};
\node (riemann) [block,below right of = mtheory] {$\Sigma$ Tiling};
\node (torus) [block,right of =riemann] {$T^2$ Brane Tiling};
\node (quiver) [block,above right of= torus] {Quiver};

\draw[<->] (toric) --  (riemann) node[midway,sloped,below]{
Mirror Symmetry};
\draw[<->] (riemann) -- (torus) node[midway,sloped,above]{Twist Map};
\draw[<->]   (torus) -- (quiver) node[midway,sloped,below]{Adjacency Map};

\end{tikzpicture}
\caption{This graph shows the flow of deriving the worldvolume theory of D3-branes probing a toric singularity inside a $CY^3$-fold.}
\label{fig:twistmap}
\end{figure}

These D6-branes intersect on $\Sigma_0$ along a graph $\Gamma$ whose intersection matrix is the adjacency matrix of the associated quiver gauge theory. This intersection graph can be ``untwisted'' in a canonical way which leads exactly to the data of the brane tiling of $T^2$ described above. See Figure \ref{fig:twistmap}. In this way, the quiver gauge theory is derived from first principles by living on the intersecting D6-branes under the mirror symmetry map of the D3-branes probing the Calabi-Yau singularity. See \cite{Feng:2005gw,Franco:2005sm,Yamazaki:2008bt} for more details. 

\section{Review of Topological Strings}
\label{app:B}
In this appendix, we will review some basics of topological string theory following \cite{Marino:2005sj}. 
Topological string theory is the reduction of string theory to the topological sector:  the part of the theory invariant under deformations of the worldsheet metric. This can be formulated from a $\CN=(2,2)$ supersymmetric non-linear sigma model $\phi:\Sigma \to X$ where we additionally include a sum over the worldsheet topologies in the path integral, thereby coupling to 2D quantum gravity. 
We then make this theory topological by performing a \emph{topological twist}. This projects onto the diagonal component of the 2D Lorentz (or rather euclidean rotation in euclidean string theory) $SO(2)_E$ times some worldsheet $U(1)$ $R$-type symmetry. Since the $SO(2)_E$ representations determine the spin of the different fields, this means that twisting by a $U(1)_R$ symmetry shifts the spins of the particles by their $U(1)_R$ charge. 

Let us see now in some more detail the construction of topological string theory.
The theory of a general $\CN=(2,2)$ non-linear sigma model with target space given by a K\"ahler manifold is described by the action
\begin{align}\begin{split}
S=\int_\Sigma d^4\theta d^2 z~ K(\Phi^i,\bar{\Phi}^{\bar{i}})~,
\end{split}\end{align}
where the $\{\Phi^i\}$ are chiral superfields with lowest component $\phi^i$ and 
\be
G_{i\bar{j}}=\frac{\partial^2K}{\partial \phi^i \partial \phi^{\bar{j}}}~,
\ee
is the K\"aler metric of the target manifold. Writing out the components of the chiral superfield as $\{\phi^i, \psi^i_{A},F^i\}$ we can write the action as
\begin{align}\begin{split}
S=-\int_\Sigma &d^2 z\Big\{G_{i\bar{j}}\big(D_z\phi^i D_{\bar{z}}\bar{\phi}^{\bar{j}}+i \epsilon^{AB}\bar{\psi}^{\bar{j}}_A D_z \psi^i_B\\&
-(F^i +\epsilon^{AB}\Gamma^i_{jk}\psi^j_A\psi^k_B)(\bar{F}^{\bar{j}}+\epsilon_{AB}\Gamma^{\bar{j}}_{\bar{k}\bar{\ell}}\bar{\psi}^{\bar{k}}_A\bar{\psi}^{\bar{j}}_B)\big)
-\epsilon^{AB}\epsilon^{CD}R_{i\bar{j}k\bar{\ell}}\psi^i_A \psi^k_B \psi^{\bar{j}}_C\psi^{\bar{\ell}}_D\Big\}~.
\end{split}\end{align}
This action is supersymmetric under the transformations
\be
\delta \phi^i =\epsilon^{AB}\eta_B \psi^i_A\quad,\qquad \delta \psi^i_A=-2i\epsilon_{AB} \bar\eta^BD_z \phi^i +\eta_AF^i~.
\ee
While this is defined for general Calabi-Yau manifolds, we will restrict to the case of a Calabi-Yau 3-fold for the remainder of this review. 

Now to make the theory topological, we must perform the topological twist. There are two choices of $R$-symmetry for this procedure: $U(1)_A$ axial symmetry and $U(1)_V$ vector symmetry. Recall that the classical bosonic symmetry of the worldsheet theory respects $SO(2)_E\times U(1)_L\times U(1)_R$ where $SO(2)_E$ is the Euclidean symmetry and $U(1)_{L,R}$ are the left and right chiral symmetries respectively. The $U(1)_{L,R}$ symmetries are often combined into vector and axial symmetries by changing basis of the generators
\begin{align}\begin{split}
&U(1)_L\times U(1)_R\cong U(1)_V\times U(1)_A\text{:}\quad F_V=F_L+F_R\quad,\quad F_A=F_L-F_R~,
\end{split}\end{align}
where $F_{L,R}$ are the generators of the Lie algebra for $U(1)_{L,R}$ respectively. Similarly we can topologically twist by projecting
\begin{align}\begin{split}
\text{A-twist: }SO(2)_E\times U(1)_V\to U(1)_{E,A}\qquad J_{E,A}=J-F_V~,\\
\text{B-twist: }SO(2)_E\times U(1)_A\to U(1)_{E,B}\qquad J_{E,B}=J+F_A~,
\end{split}\end{align}
where $J_{E,\_}$ is the generator of $U(1)_{E,\_}$. Since we are twisting the worldsheet Euclidean group by a chiral symmetry, we shift the spins of the fields of the model by their charge under $F_{V},F_A$ so that the fields are all bosonic (although they may be Grassmann fields). The theories resulting from these topological twists are referred to as the A- and B- models respectively. 

Topologically twisting, in addition to making fields bosonic, changes the spin of the supercharge operators. This allows us to define a pair of scalar supercharges for each model called the \emph{topological supercharge}
\be
\text{A-twist: }\CQ=Q_{+,+}+Q_{-,-}\quad,\qquad \text{B-twist: }\CQ=Q_{+,-}+Q_{-,-}~. 
\ee
Now since there are no bosonic symmetry currents, the supercharges must be nilpotent 
\be
\{\CQ,\CQ\}=0~.
\ee
In the case of the topological A-model we can write the action as
\begin{align}\begin{split}
S_A=\int_\Sigma d^2z \sqrt{g}\Big\{G_{i\bar{j}}\big(g^{\mu\nu}\partial_\mu \phi^i\partial_\nu \phi^{\bar{j}}+\frac{i \epsilon^{\mu \nu}}{\sqrt{g}} \partial_\mu \phi^i \partial_\nu \phi^{\bar{j}}-g^{\mu\nu}\rho^i_\mu D_\nu \chi^{\bar{j}}\\
-g^{\mu\nu}\rho_\mu^{\bar{j}}D_\nu \chi^i-\half g^{\mu\nu}F_\mu^iF_\nu^{\bar{j}}\big)+\half R_{\bar{i}j\bar{k}\ell}\rho_\mu^{\bar{i}}\rho_\nu^j\chi^{\bar{k}}\chi^\ell\Big\}~,
\end{split}\end{align}
where have made the identification
\begin{align}\begin{split}\label{eq:twistfields}
\chi^i=\psi^i_{+,+}\quad,\qquad \chi^{\bar{i}}=\psi_{-,-}^{\bar{i}}~,\\
\rho_{\bar{z}}^i=\psi^i_{-,+}\quad,\qquad \rho_z^{\bar{i}}=\psi_{+,-}^{\bar{i}}~.
\end{split}\end{align}
Now the supersymmetry transformations take the form of 
\begin{align}\begin{split}\label{eq:Asusy}
&[\CQ,\phi^i]=\chi^i\quad,\qquad \{\CQ,\chi^i\}=0~,\\
&\{\CQ,\rho_{\bar{z}}^i\}=2\partial_{\bar{z}}\phi^i-F_{\bar{z}}^i-\Gamma^{i}_{jk}\chi^j\rho_{\bar{z}}^k~,\\
&[\CQ, F_{\bar{z}}^i]=2D_{\bar{z}}\chi^i-\Gamma^i_{jk}\chi^j F_{\bar{z}}^k+R^i_{~k\bar{j}\ell}\chi^k\chi^{\bar{j}}\rho_{\bar{z}}^\ell~.
\end{split}\end{align}
Note that since the $\chi^i$ variables are of zero spin, but are grassmann variables, they can be identified with differential forms -- and their products with general wedge product of 1-forms on the target space $X$. This is related to the fact that since $CY^3$-folds have a trivial canonical bundle, we can identify the chiral spin bundle with the bundle of holomorphic differential forms and hence the non-chiral spin bundle with the cotangent bundle. In this way, we can identify $\CQ$ with the de Rahm differential operator $d$. 

Similarly the B-model can be written as
\begin{align}\begin{split}
S_B=\int_\Sigma d^2z\Big\{G_{i\bar{j}}\big(\partial_z\phi^i \partial_{\bar{z}}\phi^{\bar{j}}+\partial_{\bar{z}}\phi^i \partial_z\phi^{\bar{j}}\big)-\rho_z^i(G_{i\bar{j}}D_{\bar{z}}\eta^{\bar{j}}+D_{\bar{z}}\theta_i)\\
-\rho_{\bar{z}}^i(G_{i\bar{j}}D_z\eta^{\bar{j}}-D_{z}\theta_i)-R^i_{~j\bar{\ell} k}\eta^{\bar{\ell}}\rho_z^j\rho_{\bar{z}}^k\theta_i-G_{i\bar{j}}F^iF^{\bar{j}}\Big\}~,
\end{split}\end{align}
where we have again defined the fields as in \ref{eq:twistfields} with the additional identification
\be
\eta^{\bar{i}}=\chi^{\bar{i}}+\bar{\chi}^{\bar{i}}\quad,\qquad \theta_i=G_{i\bar{j}}(\chi^{\bar{j}}-\bar{\chi}^{\bar{j}})~.
\ee
Using the B-twisted supercharge, these fields satisfy the supersymmetry relations
\begin{align}\begin{split}
[\CQ,\phi^i]=0\quad,\qquad [\CQ,\phi^{\bar{i}}]=\eta^{\bar{i}}~,\\
\{\CQ,\eta^{\bar{i}}\}=0\quad,\qquad \{\CQ,\theta_i\}=G_{i\bar{j}}F^{\bar{j}}~,\\
\{\CQ,\rho_z^i\}=\partial_z \phi^i\quad,\qquad \{\CQ,\rho_{\bar{z}}^i\}=\partial_{\bar{z}}\phi^i \\
\{\CQ,F^i\}=D_z\rho_{\bar{z}}^i-D_{\bar{z}}\rho_z^i+R^i_{~j\bar{\ell}k}\eta^{\bar{\ell}}\rho_z^j\rho_{\bar{z}}^k\\
\{\CQ,F^{\bar{i}}\}=-\Gamma^{\bar{i}}_{\bar{j}\bar{k}}\eta^{\bar{j}}F^{\bar{k}}~,
\end{split}\end{align}
Again since we have that $\chi^i$ (and hence $\eta^i,\theta^i$) can be identified with 1-forms on the target space $X$. Therefore from the supersymmetry relations relating $\phi^i$ and $\eta^i$, we can associate the supercharge operator $\CQ_B=\bar\partial$.


This general analysis uses what is called the Mathai-Quillen formulation of a topological non-linear sigma model \cite{Mathai:1986tc}. The above analysis tells us that the A-model deals with the de Rahm cohomology of the Calabi-Yau manifold whereas the B-model deals with Dolbeault cohomology and hence the A-model is invariant under complex structure deformations while the B-model is invariant under K\"ahler deformations. Since mirror symmetry exchanges complex structure and K\"ahler moduli, one would (correctly) expect that the A- and B- models are mirror dual. For more details see \cite{Hori:2003ic,Marino:2005sj,Neitzke:2004ni}.

Thus far we have described some of the general features of these twisted quantum field theories. In order to show that these theories are topological we need to show that the expectation value of supersymmetric operators is invariant under a deformation of the metric. 

Note that the action of these topologically twisted theories can be written as
\be
S_{TFT}=\{\CQ,V\}~,
\ee
where 
\begin{align}\begin{split}
V_A&=\half\int_\Sigma d^2z\sqrt{g}g^{\mu\nu}G_{i\bar{j}}\left[\half \rho_\mu^i F_\nu^{\bar{j}}+\half \rho_\mu^{\bar{j}}F_\nu^i+(\rho_\mu^i \partial_\nu \phi^{\bar{j}}+\rho_\mu^{\bar{j}}\partial_\nu \phi^i)\right]~,\\
V_B&=\int_\Sigma d^2z\sqrt{g}\left[G_{i\bar{j}}g^{\mu\nu}\rho_\mu^i \partial_\nu \phi^{\bar{j}}-F^i\theta_i\right]~,
\end{split}\end{align}
for the A- and B-model respectively. Since the stress energy tensor is defined by
\be
T_{\mu\nu}=\frac{\delta S}{\delta g^{\mu\nu}}~,
\ee
we have that in these cases
\be
T_{\mu\nu}=\{\CQ,G_{\mu\nu}\}~,
\ee
for some tensor $G_{\mu\nu}$. 

Now consider the expectation value of a collection of operators $\{\CO_i\}$ such that 
\be
\{\CQ,\CO_i\}=0\qquad\forall~i~.
\ee
By use of the Ward identities, we have that the variation of their expectation value with respect to the worldsheet metric
\be
\frac{\delta}{\delta g^{\mu\nu}}\langle \CO_1 ... \CO_n\rangle=\langle \CO_1... \CO_n \delta G_{\mu\nu}\rangle=\langle \delta(\CO_1... \CO_nG_{\mu\nu})\rangle=0
\ee
Now since the action is $\delta$-exact, we have that this expectation value vanishes order by order by integrating by parts in field space\footnote{We are ignoring the subtleties coming from boundary contributions in field space. See \cite{Witten:1988ze,Witten:1988xj,Witten:1988xi} for more details.}. Therefore, since the expectation value of a product of any (supersymmetric) operators is invariant under variations of the metric, these theories are topological.


Note that this similar, but distinct to the idea of localization. In the case of localization, we again have that 
\be
S=\int_{M}d^dx~\{Q,V\}~,
\ee
for some potential function $V$ and $Q$ a supercharge and we want to compute the expectation value of some collection of supersymmetric operators $\{\CO_i\}$ where $\delta \CO_i=0$ $\forall i$. In this case we can integrate by parts in field space so that $Q$ acts on the operators $\CO_i$. In this case the expectation value reduces to the path integral integral over the zero-locus of $V$ in field space -- hence \emph{localizing} to the zeros of $V$. See \cite{Pestun:2016zxk} for more details on localization.

\subsection{Correlation Functions in Topological Field Theory}

So far, we have only introduced topological field theories since we have not yet implemented a sum over worldsheet topologies. However, even the topological field theory expectation values give insight into understanding the associated topological string theory since they represent the tree level contribution. As in quantum field theory, understanding the anomalies and ghost structure of the theory is crucial to our understanding of these theories. In these topologically twisted theories we can interpret bosonic, Grassmann supercharge $\CQ$ as the BRST charge. This means that we can associate the ghost number of a given operator with the level of the operator as a differential form. \\

\noindent\textbf{\underline{A-model}:} Recall that in the A-model the different fields can be associated with $p$-forms in $H^p(X)$. From the supersymmetry transformations \ref{eq:Asusy}, we can see that the $\chi^i$ should be associated with primitive 1-forms (ghost charge $+1$) and the $\rho_z^i$ should be associated with primitive tangent vectors (ghost charge $-1$). Therefore, we should identify generic $\CQ$-closed operators of the form
\be
\CO_\alpha=\alpha_{i_1,...,i_p}\chi^{i_1}...\chi^{i_p}~,
\ee
where $\alpha=\alpha_{i_1,...,i_p}dx^{i_1}\wedge...\wedge dx^{i_p}$ is a closed $p$-form on $X$. This theory has a $U(1)$ chiral symmetry which rotates the $\chi^i$ and $\rho_z^i$ according to their ghost charge. However, this symmetry can be anomalous because of the non-trivial topology of $X$ and $\Sigma$. These fields couple differently to the worldsheet topology because $\rho_\mu^i$ is a vector and $\chi^i$ is a scalar on the worldsheet. 

The calculation of the anomaly leads to the selection rule that $\langle \CO_{\alpha_1}...\CO_{\alpha_k}\rangle\equiv0$ unless
\be
\sum_{i=1}^k \text{ deg }\CO_{\alpha_i}=2d(1-g)+2\int_\Sigma \phi^\ast (c_1(X))~,
\ee
where $g$ is the genus of $\Sigma$ and $d=$dim$_\IC X$. In our case, $X$ is a $CY^3$-fold so that the superselection rule becomes
\be
\sum_{i=1}^k \text{ deg }\CO_{\alpha_i}=6-6g~.
\ee
This tells us that only the case of $\Sigma=S^2$ has non-trivial expectation values. 

Let us consider the expectation value of three rank 2 operators $\langle \CO_{\alpha_1}\CO_{\alpha_2}\CO_{\alpha_3}\rangle$ in the A-model with $\Sigma=S^2$. Since we have that the action is $\CQ$-exact, we have that the expectation value will localize to zeros of $V_A$, each of which represents a different instanton sector\footnote{We can see this more explicitly by using the Fubini-study metric of the 2-sphere $ds^2=\frac{dz d\bar{z}}{(1+|z|^2)^2}$.}. Because  the zeros of $V_A$ are given exactly by the holomorphic maps
\be
\phi^i:\Sigma\to X\quad,\qquad \bar\partial \phi^i=0~.
\ee
the different instanton sectors are classified by holomorphic 2-cycles in $\sigma\in H_2(X;\IZ)$ which are the image of the genus 0 worldsheet. Because of the selection rules above, this means that the computation of the partition function will essentially reduce to the triple intersection of the associated 2-cycles. To be explicit, if we pick a basis of 2-cycles $[S_i]\in H_2(X;\IZ)$, where $i=1,...,b_2(X)$, then we have that 
\begin{align}\begin{split}
\langle \CO_{\alpha_1}\CO_{\alpha_2}\CO_{\alpha_3}\rangle_{\{t_i\}}=\#(\alpha_1\cap\alpha_2\cap&\alpha_3)+\sum_{\beta\in H_2(X;\IZ)}I_{\beta}(\alpha_i)Q^\beta~,\\
I_\beta(\alpha_i)=N_{0,\beta} \int_\beta \alpha_1\int_\beta\alpha_2\int_\beta\alpha_3&\quad,\qquad \alpha_i\in H^2(X;\IZ)~,
\end{split}\end{align}
and
\be
Q^\beta=\prod_{i=1}^{b_2(X)}e^{n_i t_i}\quad,\qquad \beta=\sum_{i=1}^{b_2(X)}n_i [S_i]\quad,\qquad t_i=\int_{S_i}\omega~,
\ee
where $\omega$ is the (complexified) K\"ahler class\footnote{The complexified K\"ahler class is a shift of the ``normal'' K\"aher class by an additional $B\in H^2(X;\IZ)$ which represents coupling the theory to a non-trivial B-field.}. These $N_{0,\beta}$ are an example of \emph{Gromov-Witten} invariants. In fact, these are what determine the prepotential (that is the free action from the genus zero worldsheet) 
\be
F_0(t)=\sum_\beta N_{0,\beta}Q^\beta~.
\ee

\noindent\textbf{\underline{B-model}:} Now consider the B-model. Now instead of de Rahm cohomology, the B-model relies on Doubeault cohomology. So $\CQ$-closed operators are generically of the form
\be
\CO_{\alpha}=\alpha^{j_1,...,j_q}_{\bar{i}_1,...,\bar{i}_p}\eta^{\bar{i}_1}...\eta^{\bar{i}_p}\theta_{j_1}...\theta_{j_q}~,
\ee 
where 
\be
\alpha=\alpha^{j_1,...,j_q}_{\bar{i}_1,...,\bar{i}_p}dx^{\bar{i}_1}\wedge...\wedge dx^{\bar{i}_p}\frac{\partial}{\partial x^{j_1}}\wedge...\wedge\frac{\partial}{\partial x^{j_q}}\in H^p_{\bar{\partial}}(X,\Lambda^qTX)~,
\ee
or alternatively which is canonically isomorphic to 
\be
\alpha\to \hat\alpha=\alpha^{j_1,...,j_q}_{\bar{i}_1,...,\bar{i}_p}d\bar{z}^{\bar{i}_1}\wedge...\wedge d\bar{z}^{\bar{i}_p}\wedge dz_{j_1}\wedge...\wedge dz_{j_q}\in H^{(p,q)}(X;\IZ)~.
\ee
 Again we have an anomalous chiral symmetry which rotates the $\eta^i,\theta_i$ fields according to their ghost charge. However, this time since they have the same worldsheet and spacetime properties, we have a selection rule that only couples to the topology of $\Sigma$ which constrains the expectation value of $\langle \CO_{\alpha_1}...\CO_{\alpha_n}\rangle$ to vanish identically unless
 \be
 \sum_i p_{\alpha_i}=\sum_i q_{\alpha_i}=d(1-g)~,
 \ee
where $(p_\alpha,q_\alpha)$ are the holomorphic and anti-holomorphic components of $\alpha$, $g$ is the genus of $\Sigma$, and $d=$dim$_\IC X$. 

Now let us consider the expectation value of three $\CQ$-exact rank (1,1) operators $\langle \CO_{\alpha_1}\CO_{\alpha_2}\CO_{\alpha_3}\rangle$ in the B-model.  Again we have that the expectation value localizes to the zeros of $V_B$. The instanton sectors are described by the constant maps
\be
\phi^i:\Sigma\to X\quad,\qquad d\phi^i=0\quad,\quad \forall~i~.
\ee
This means that there are no non-trivial instanton sectors in the topological B-model.

Since we are on a CY 3-fold, there is a unique $(3,0)$-form $\Omega$ which induces an isomorphism between
\be
\Omega:H^d_{\bar\partial}(X,\Lambda^qX)\to H^{(d-q,p)}(X,\IZ)~,
\ee
by contracting with the free indices of $\Omega$. This means that we can write the expectation value 
\be
\langle \CO_{\alpha_1}\CO_{\alpha_2}\CO_{\alpha_3}\rangle=\int_X \alpha_1\wedge \alpha_2\wedge \alpha_3=\int_X (\alpha_1)_{\bar{j}_1}^{i_1}(\alpha_2)_{\bar{j}_2}^{i_2}(\alpha_3)_{\bar{j}_3}^{i_3}\Omega_{i_1i_2i_3} d\bar{z}^{\bar{j}_1}\wedge d\bar{z}^{\bar{j}_2}\wedge d\bar{z}^{\bar{j}_3}\wedge \Omega~.
\ee
This rewriting allows us to encode the data purely in terms of the topological data encoded by the unique holomorphic 3-form $\Omega$. 

Fix a complex structure $J\in \CM_{c.s.}$ for $X$ and choose a local symplectic basis for $H_3(X)$ in a local patch in $\CM_{c.s.}$ containing $J$ which we will denote by $(A_a,B^b)$ with $a,b=0,...,h^{2,1}$ such that 
\be
A_a\cap B^b=\delta_a^{~b}\quad,\qquad A_a\cap A_b=B^a\cap B^b=0~.
\ee
Now define the periods of the CY manifold
\be
z_a=\int_{A_a}\Omega\quad,\qquad F^a=\int_{B^a}\Omega~.
\ee

Since $H^{(2,1)}(X,\IZ)$ parametrizes the complex moduli $\CM_{c.s.}$ of the theory, without loss of generality, we can take the $z_a$ as projective coordinates parametrizing $\CM_{c.s.}$. This means that the vector space of $F^a$ form a fiber of a vector bundle over $\CM_{c.s.}$. Therefore, they can individually be thought as functions of the $z^a$. 

Using this we can define the generating function which is called the \emph{prepotential}
\be
F_0=\half \sum_a z_aF^a~.
\ee
Since $z_a$ are projective coordinates and the function $F_0$, which is the genus zero contribution to the free energy, is a homogeneous polynomial in terms of these coordinates (a rescaling of the $z_a$ is associated with the rescaling of $\Omega$), the physical data is encoded in the scale invariant projection of $F_0$. Now the expectation value $\langle \CO_{\alpha_1}\CO_{\alpha_2}\CO_{\alpha_3}\rangle$ where $deg~\CO_{\alpha_i}=(1,1)$, which is given by the triple intersection of the associated 2-cycles $\#(\hat\alpha_1\cap\hat\alpha_2\cap\hat\alpha_3)$. We can write this as
\begin{align}\begin{split}
\langle \CO_{\alpha_1}\CO_{\alpha_2}\CO_{\alpha_3}\rangle=&\sum_{a,b,c,}n_am_b\ell_c\frac{\partial^3 F_0}{\partial z_a\partial z_b\partial z_c}~,\\
\hat\alpha_1=\sum_a n_a[A_a]\quad,\qquad \hat\alpha_2=&\sum_bm_b[A_b]\quad,\qquad \hat\alpha_3=\sum_c \ell_c[A_c]~.
\end{split}\end{align}

In the literature we often use the coordinates $t_a$ instead of the $z_a$ to denote the special projective coordinates which have removed the scaling dependence of the $z_a$. These are only defined patchwise on $\CM_{c.s.}$.

\subsection{Topological Strings and Mirror Symmetry}

In order to construct topological string theory we have to couple our theory to 2D topological gravity. This means that in the path integral we have to include a sum over the different worldsheet topologies $\Sigma_g$ weighted by $g_s^{-\chi(\Sigma_g)}=g_s^{2g-2}$. Now we see that the free action $F=\log Z$ will generically have the form
\be
F(t)=\sum_{g=0}^\infty g_s^{2g-2} F_g(t)~.
\ee
As it turns out here $F_0(t)$ is exactly the $F_0$ prepotential computed above in the A- and B-twisted topological field theories. These higher genus terms are generically very difficult to compute generally relying on mirror symmetry and the holomorphic anomaly equations  \cite{Bershadsky:1993ta,Bershadsky:1993cx}. The computation of these terms is still an active field of research. 

A salient difference between the A- and B- topological strings is that the A-model depends only on K\"ahler structure whereas the B-model depends on the complex structure. Since we have that mirror symmetry exchanges complex and K\"ahler geometry, mirror symmetry exchanges the A- and B-models. This turns out to be a direct consequence of the mirror symmetry between Type IIA and Type IIB.

\subsection{Relation to 4D $\mathcal{N}=2$ Theories}
\label{sec:App4D}

It is well known that compactifying Type IIA or Type IIB string theory on a $CY^3$-fold produces a four-dimensional $\CN=2$ theory. In the limit that the volume of the $CY^3$ manifold becomes infinitely large (or is non-compact) and $g_s,\ell_s\to 0$ while keeping $g_{YM}$ constant, we decouple from gravity producing a $\CN=2$ SUSY QFT. Many SUSY operators in the 4D $\CN=2$ theory can be computed in the associated topological string theory given by taking topological string theory on the associated $CY^3$ manifold as suggested in the previous section. For example, the partition function of the topological string computes exactly that of the 4D $\CN=2$ SUSY QFT. Many of the SUSY operators in these theories can be realized as BPS branes in the full string theory which can be interpreted clearly in the topological string picture, giving us the ability to calculate their expectation value using the techniques of topological string theory.

%
%

Recall that in the B-model the prepotential is given in terms of the periods of the holomorphic 3-form $\Omega$ over the 3-cycles in $H_3(Y,\IZ)$
\be
F_0=\frac{1}{2}z_iF^i\quad,\qquad z_i=\int_{A^i} \Omega\quad,\qquad F^i=\int_{B_i}\Omega~.
\ee
This is of course reminiscent (and is in fact exactly the generalization of) the relation between Sieberg-Witten theory and the instanton partition function calculation of Nekrasov.

In fact, the topological string computes exactly the low energy effective action of the vector multiplet for the 4D $\CN=2$ theory from compactifying Type II string theory on the $CY^3$ manifold Y. Specifically we find that if $F_g(t)$ are the terms in the topological string free energy, then the low energy effective theory will go as 
\be
S_{EFT}\sim \int d^4xd^4\theta\text{ }\CW^{2g}F_g(a^i)\sim \int d^4F_g(a^i)(R^2_+F_+^{2g-2})~,
\ee
for the case where $g\neq 0$, $a^i$ are the vector multiplet fields, $\CW$ is the $\CN=2$ Weyl multiplet\footnote{That is the multiplet with highest component given by the field strength of the gravitational multiplet: $R_+$.} $R_+^2$ is the self-contraction of the self-dual part of the Riemann tensor and $F_+$ is the self-dual part of the field strength of the graviphoton. See \cite{Hori:2003ic} for more details. 


For the case $g=0$, we find that the effective action is given by:
\be
\int d^4 x(\partial_i\partial_jF_0)F_i^+\wedge F_j^+=\int d^4x\text{ }\tau_{ij}\text{ }F_i^+\wedge F_j^+~,
\ee
which is exactly the same form as the Seiberg-Witten low energy effective action \cite{Seiberg:1994rs,Seiberg:1994aj}. Hence the genus-0 term gives us exactly the prepotential for the 4D $\CN=2$ gauge theory \cite{Dijkgraaf:2009pc}.

If we consider the topological string theory on a CY-3fold of the form \cite{Klemm:1996bj,Dijkgraaf:2002fc,Dijkgraaf:2009pc,Cheng:2010yw}
\be
\{f(u,v,y,x)=uv+y^2+W'(x)^2=0\}\subset \IC^4~,
\ee
then the holomorphic 3-form is of the form
\be
\Omega=\frac{du\wedge dy\wedge dx}{\partial_v f}=\frac{du}{u}\wedge dy\wedge dx~.
\ee
In this case, the period integrals reduce to integrals localized at $u=0$ by the nature of the residue theorem. That is to say, the period integrals reduce to integrals over a disc whose boundary is a 1-cycle $\gamma\subset\Sigma$ where 
\be
\Sigma=\{y^2+W'(x)^2=0\}~,
\ee
such that the period integral becomes
\be
\int_{\sigma\subset X}\Omega=\int_D dy\wedge dx=\int_{\gamma\subset \Sigma}y dx~.
\ee
Here we can clearly identify the pair of Seiberg-Witten data $(y dx,\Sigma)=(\lambda_{SW},\Sigma_{SW})$ from first principles \cite{Aganagic:2011mi}. There is also a way to use this picture and more advanced topics in the topological B-model to derive the work of \cite{Nekrasov:2009rc} which relates theories of class $\CS$ to integrable models.

\bigskip

\bigskip

\nocite{*}
\bibliographystyle{JHEP}
\bibliography{LandscapeSwamplandCorner}{}

\end{document}